\documentclass[3p]{elsarticle}
\UseRawInputEncoding
\usepackage{lineno,hyperref}
\usepackage{graphicx}
\usepackage{amsmath}
\usepackage{makecell}
\usepackage{booktabs}
\usepackage{caption}
\usepackage{tikz}
\usetikzlibrary{arrows.meta,positioning}
\hypersetup{hidelinks}

\journal{arXiv}
\begin{document}

\begin{frontmatter}

\title{From a Hierarchy of Stochastic Differential Equations to a Hierarchy of Generalized Beta Distributions}

\author[mymainaddress]{Siqi Shao}
\author[mymainaddress]{R. A. Serota\fnref{myfootnote}}
\fntext[myfootnote]{serota@ucmail.uc.edu}

\address[mymainaddress]{Department of Physics, University of Cincinnati, Cincinnati, Ohio 45221-0011}

\begin{abstract}

We introduce a mean-reverting stochastic differential equation with a three-component stochastic term and show that it generates a hierarchy of steady-state (stationary) distributions. At the top level, the hierarchy is described by a modified-Beta distribution, while one- and two-parameter reductions produce compact-support, power-law-tailed, and exponential-type limiting families within a single stochastic framework. 

We then construct two generalized extensions of this hierarchy. In the first, the power transformation is applied directly at the level of the stochastic differential equation; in the second, the same transformation is applied only after the stationary modified-Beta hierarchy has been obtained. While these two procedures agree on important lower branches and limiting cases they generally differ at the top level. The generalized hierarchy is therefore not unique: nonlinear transformation and stationary-state reduction do not commute.

For both routes, we derive the probability density and cumulative distribution functions, express their parameters in terms of the underlying stochastic dynamics, clarify the relations among their limiting cases, and compare the resulting families with the traditional Generalized Beta framework.

\end{abstract}


\end{frontmatter}

\section{Introduction}

The Generalized Beta (GB) family of distributions occupies a special place in statistics because it unifies distributions with finite support, power-law tails, and exponential-type suppression within a single flexible framework. It is widely used in econometrics for study of income and wealth distributions \cite{mcdonald1984some,mcdonald1995generalization,sepanski2007family,chotikapanich2008modelling,sarabia2014explicit,chotikapanich2018using,gomez2018family,biewen2018econometrics} and it finds applications in such diverse fields as polymer physics \cite{sepplinen2012scaling,thiery2014loggamma,grange2017loggamma} and cognitive psychology \cite{dashti2020modeling}. In parallel, a number of dynamical models in econophysics \cite{chakrabarti2013econophysics} and mathematical finance, many based on stochastic differential equations (SDE) formalism \cite{risken1996fokker,jacobs2010stochastic}, were developed that produced individual members of GB family of distributions for quantities such as wealth and stochastic variance. Examples of the latter include Inverse Gamma (IGa) \cite{praetz1972distribution, nelson1990arch,bouchaud2000wealth,fuentes2009universal,liu2019distributions}; Generalized Inverse Gamma (GIGa) \cite{ma2013distribution}; Beta Prime (BP), also known as B2 \cite{dashti2020stochastic,dashti2021combined}; and Gamma (Ga) \cite{cox1985theory,heston1993closed,dragulescu2002probability}.

Initial ground-breaking attempt to systemize branches of GB family of distributions in terms of steady-state (stationary) distributions of SDE was undertaken in \cite{hertzler2003classical} but did not find deserved recognition. It was subsequently generalized and greatly simplified in \cite{liu2023rethinking}. This gives GB distributions landscape a dynamical interpretation and relates their distributional parameters to underlying stochastic mechanisms. As a result, steady-state distributions that were previously derived in a case-by-case manner became possible to view within a single stochastic scheme that generates an entire hierarchy of related steady-state distributions and clarifies the relations among their limiting forms. This hierarchy compliments and, in certain ways, supplants the hierarchy in \cite{mcdonald1995generalization} which made its way to Wikipedia \cite{gbdistributiontree}.

Our starting point is a mean-reverting stochastic differential equation whose stochastic term contains three independent parameters. Its stationary solution is a modified-Beta (mB) distribution \citep{liu2023rethinking}, and controlled parameter reductions generate the lower branches and terminal limits of the hierarchy. In this way, compact-support, heavy-tailed, and exponential-type families appear as dynamically connected members of a single stochastic construct rather than as separate distributions introduced case by case.

An important question addressed here is how this hierarchy should be generalized under the power transformation \(y=x^\alpha\). There are two natural but inequivalent ways to proceed. In the \emph{dynamical route}, one transforms the stochastic differential equation first and then derives the stationary distribution of the transformed dynamics. In the \emph{distributional route}, one first derives the stationary hierarchy in the Beta sector and then pushes those stationary distributions forward under the same transformation. The important structural result of the paper is that the two procedures share the same reduction pattern but generally do not produce identical generalized
distributions. They coincide with the original Beta-sector hierarchy when \(\alpha=1\), while for \(\alpha\neq 1\) the endpoint exponents and normalization
constants become route-dependent. The generalized hierarchy is therefore not unique: nonlinear transformation and stationary-state reduction do not commute.

This point is especially relevant when compared with the traditional Generalized Beta family, which is usually introduced directly at the level of probability density functions \cite{mcdonald1995generalization,cordeiro2009new,alexander2012generalized,alzaatrech2013newmethod,liu2023rethinking}. By contrast, the present framework starts from stochastic dynamics and derives the hierarchy of distributions as a hierarchy of stationary states. It therefore provides not only a common dynamical origin for distributions with finite support, power-law tails, and exponential-type suppression, but also two distinct modified generalized-Beta top-level extensions associated with that origin.

For clarity, we refer to the hierarchy generated directly by the master stochastic differential equation as the \emph{Beta sector}. Its two generalized extensions are called the \emph{dynamical route}, obtained by transforming the stochastic differential equation itself, and the \emph{distributional route}, obtained by transforming the stationary distributions.

The paper is organized as follows. In Sec.~2, we introduce the mean-reverting stochastic differential equation and the modified-Beta hierarchy generated by it. In Sec.~3, we develop the generalized hierarchy obtained directly from the transformed stochastic equation (the \emph{dynamical route}) and derive the corresponding probability density and cumulative distribution functions. In Sec.~4, we present the hierarchy obtained by power-transforming the modified-Beta family (the \emph{distributional route}) and compare it with the direct generalized construction. In Sec.~5, we discuss the relation between these hierarchies and the traditional Generalized Beta family. In Sec.~6, we conclude with a discussion and outlook.

\section{Hierarchy Generated by the Master Stochastic Differential Equation}

We begin with a mean-reverting stochastic differential equation whose
stochastic term contains three independent contributions. The purpose of this
construction is to place, within a single dynamical framework,
finite-support distributions, heavy-tail distributions, and their simpler
limiting forms. Rather than introducing each distribution separately, we
start from one master equation and then obtain the hierarchy by controlled
parameter reductions.

We consider stochastic process for positively defined variable $x$
\begin{equation}
dx
=
-\gamma(x-\theta)\,dt
+
\sqrt{
\left(1-\kappa_1^2 x\right)
\left(\kappa^2 x+\kappa_2^2 x^2\right)
}\,dW_t ,
\label{eq:master_sde_kappa}
\end{equation}
where \(dW_t\sim \mathcal N(0,dt)\) is the Wiener increment,
\(\gamma>0\) is the mean-reversion rate, and \(\theta>0\) is the
mean-reversion scale. The latter means that, sans stochastic term, $x$ relaxes to $\theta$ 
from any initial value over time $\sim \gamma^{-1}$. 
It also means that, with the full SDE (\ref{eq:master_sde_kappa}), $\langle x \rangle=\theta$ 
(mean reversion) once the distribution of $x$ relaxes to its stationary state. 
Parameter \(\kappa\) controls the linear-in-\(x\)
stochastic contribution, \(\kappa_1\) controls the finite-support cutoff, and
\(\kappa_2\) controls the multiplicative quadratic contribution responsible
for the heavy-tail sector.

It is often convenient to introduce two scale parameters
\begin{equation}
\beta_1=\frac{1}{\kappa_1^2},
\qquad
\beta_2=\frac{\kappa^2}{\kappa_2^2},
\label{eq:beta_map}
\end{equation}
in terms of which Eq.~\eqref{eq:master_sde_kappa} can be rewritten as
\begin{equation}
dx
=
-\gamma(x-\theta)\,dt
+
\sqrt{
\kappa^2 x
\left(1-\frac{x}{\beta_1}\right)
\left(1+\frac{x}{\beta_2}\right)
}\,dW_t .
\label{eq:master_sde}
\end{equation}
The factor \((1-x/\beta_1)\) imposes a finite upper endpoint at
\(x=\beta_1\) (finite-support structure), while the factor \((1+x/\beta_2)\) governs the crossover
toward the heavy-tail regime. In this way, the master equation interpolates
between compact-support and unbounded-support sectors within a single
stochastic framework.

The key advantage of Eq.~\eqref{eq:master_sde_kappa} is that it generates
a hierarchy of steady-state distributions by controlled parameter reductions \citep{liu2023rethinking}:
\begin{itemize}
\item the full equation produces the top-level mB distribution;
\item setting \(\kappa=0\) yields the finite-support inverse-Gamma (IGa1) branch \footnote{Due to the difference in parametrization this branch was omitted in \citep{liu2023rethinking}};
\item setting \(\kappa_1=0\) yields the unbounded modified Beta-prime (mBP) branch;
\item setting \(\kappa_2=0\) yields the compact-support Beta type-I (mB1) branch.
\end{itemize}
Thus, the hierarchy is not imposed externally -- it is encoded directly in the
structure of the stochastic term.

At the level of  one-parameter reductions, the hierarchy is
therefore summarized schematically as follows\footnote{Since mB1 coincides with traditional B1 and mBP is virtually identical to BP (see \cite{dashti2021combined}, Eq.(15) in footnote of \cite{liu2023rethinking}, and Sec. \ref{traditional} below), prefix $m$ can be omitted in mB1 and mBP or used interchangeably.}:
\begin{equation}
mB
\;\xrightarrow{\;\kappa=0\;}\;
IGa1,
\qquad
mB
\;\xrightarrow{\;\kappa_1=0\;}\;
BP/B2,
\qquad
mB
\;\xrightarrow{\;\kappa_2=0\;}\;
B1 .
\label{eq:one_param_hierarchy_intro}
\end{equation}
The two-parameter limits of these branches produce the simpler Ga and
IGa discussed below. In this manner, the familiar Beta (B),
BP, Ga, and IGa distributions are seen to be
descendants of a single stochastic ancestor rather than isolated objects.

\subsection{Steady-state distribution and parameter mapping}

We now derive the stationary distribution associated with the master
stochastic differential equation \eqref{eq:master_sde_kappa}. It is convenient to introduce positive shape parameters
\begin{equation}
p=\frac{2\gamma\theta}{\kappa^2},
\qquad
q=\frac{2\gamma(1-\kappa_1^2\theta)}{\kappa^2\kappa_1^2+\kappa_2^2}
=
\frac{2\gamma(\beta_1-\theta)}
{\kappa^2\left(1+\beta_1/\beta_2\right)}.
\label{eq:pq_map}
\end{equation}
In terms of these parameters, the stationary probability density function (PDF) can
be written as \cite{liu2023rethinking,liu2024dragon}
\begin{align}
f_{mB}(x;\beta_1,\beta_2,p,q)
&=
\frac{(p+q)\left(1+\beta_1/\beta_2\right)^{p+1}}
{\beta_1\left[p+\left(1+\beta_1/\beta_2\right)q\right]B(p,q)}
\left(\frac{x}{\beta_1}\right)^{p-1}
\left(1-\frac{x}{\beta_1}\right)^{q-1}
\left(1+\frac{x}{\beta_2}\right)^{-p-q-1} \nonumber \\
&\quad
=\frac{(p+q)\left(1+\beta_2/\beta_1\right)^{p+1}}
{\beta_2\left[q+\beta_2/\beta_1(p+q)\right]B(p,q)}
\left(\frac{x}{\beta_2}\right)^{p-1}
\left(1-\frac{x}{\beta_1}\right)^{q-1}
\left(1+\frac{x}{\beta_2}\right)^{-p-q-1},
\label{eq:mB_pdf}
\end{align}
with $0\le x\le \beta_1$. Here $B(p,q)=\Gamma(p)\Gamma(q)/\Gamma(p+q)$ 
is the beta function and $\Gamma$ is the gamma function \cite{nist2026digital}.
Normalizability of (\ref{eq:mB_pdf}) is obviously assured by the fact that both $p>0$ and $q>0$.

PDF \eqref{eq:mB_pdf} is the top-level modified-Beta density
generated by the master stochastic dynamics. The factor
$\propto x^{p-1}$, which can be scaled with either $\beta_1$ or $\beta_2$, 
controls the small-\(x\) behavior, the factor
\((1-x/\beta_1)^{q-1}\) controls the approach to the upper edge
\(x=\beta_1\), and the factor \((1+x/\beta_2)^{-p-q-1}\) governs the
crossover away from the finite-support Beta-like form toward the lower
branches of the hierarchy. In this sense, the parameters
\((p,q,\beta_1,\beta_2)\) are not arbitrary fitting constants, but compact
encodings of the mean-reversion and fluctuation structure of the underlying
stochastic equation. 

As discussed in detail below, being the top node, mB PDF \eqref{eq:mB_pdf} initiates flows 
to the lower branches of the family when one of the stochastic contributions is removed. 
Also very important is that for $\beta_2 \ll \beta_1, $ (\ref{eq:mB_pdf})
predicts that $\propto (x/\beta_2)^{p-1}$ behavior for $x\ll\beta_2$ gives way to 
power-law decay $\propto (x/\beta_2)^{-q-2}$ for $\beta_2 \ll x \ll \beta_1$ before 
termination at $\beta_1$.

For later use, it is also helpful to record the cumulative distribution
function (CDF) in a closed form. Defining
\begin{equation}
z(x)=\frac{x/\beta_1+x/\beta_2}{1+x/\beta_2},
\label{eq:z_def}
\end{equation}
the CDF can be written as
\begin{align}
F_{mB}(x;\beta_1,\beta_2,p,q)
&=
I\!\bigl(z(x);p,q\bigr)
\nonumber\\[4pt]
&\quad
+\frac{1}{B(p,q)\left[
q+\dfrac{\beta_2}{\beta_1}(p+q)
\right]}
\left(\frac{1-x/\beta_1}{1+x/\beta_2}\right)^q
\left(
\frac{(1+\beta_1/\beta_2)\,x/\beta_1}{1+x/\beta_2}
\right)^p ,
\label{eq:mB_cdf}
\end{align}
where
\begin{equation}
I(z;p,q)=\frac{B(z;p,q)}{B(p,q)}
\end{equation}
is the regularized incomplete beta function \cite{nist2026digital}.

The representation \eqref{eq:mB_cdf} is especially useful because it makes
transparent the relation between the present hierarchy and the traditional
Beta and generalized-Beta families \cite{mcdonald1995generalization}, whose CDF are also
naturally expressed in terms of incomplete beta functions, namely the first term in (\ref{eq:mB_cdf}) \cite{liu2023rethinking}. 

\subsection{One-parameter reductions: the \texorpdfstring{$\kappa=0$}{kappa=0}, \texorpdfstring{$\kappa_1=0$}{kappa1=0}, and \texorpdfstring{$\kappa_2=0$}{kappa2=0} branches}

The hierarchy generated by Eq.~\eqref{eq:master_sde_kappa} becomes especially
transparent when one of the three stochastic contributions is removed. We
present the three resulting one-parameter reductions in the order
\(\kappa=0\), \(\kappa_1=0\), and \(\kappa_2=0\), which is also the order
used in Table~\ref{table1}.

\subsubsection{Finite-support inverse-Gamma branch: the \texorpdfstring{$\kappa=0$}{kappa=0} limit}

Setting
\begin{equation}
\kappa=0
\end{equation}
removes the linear-in-\(x\) stochastic contribution while retaining the
finite-support factor and the multiplicative \(x^2\) noise. The master
equation \eqref{eq:master_sde_kappa} then reduces to
\begin{equation}
dx
=
-\gamma(x-\theta)\,dt
+
\kappa_2 x\sqrt{1-\kappa_1^2 x}\,dW_t
=
-\gamma(x-\theta)\,dt
+
\kappa_2 x\sqrt{1-\frac{x}{\beta_1}}\,dW_t ,
\label{eq:IGa1_sde}
\end{equation}
with support \(0<x<\beta_1\), where \(\beta_1=1/\kappa_1^2\).

The corresponding stationary density is
\begin{equation}
f_{IGa1}(x;\beta,\beta_1,q)
=
c_{\text{IGa1}}e^{-\beta/x}\left(\frac{x}{\beta}\right)^{-q-2}\left(1-\frac{x}{\beta_1}\right)^{q-1},
\qquad
0<x<\beta_1 ,
\label{eq:IGa1_pdf}
\end{equation}
where
\begin{equation}
\beta=\frac{2\gamma\theta}{\kappa_2^2},
\qquad
q=\frac{2\gamma(1-\kappa_1^2\theta)}{\kappa_2^2}
=\frac{2\gamma(\beta_1-\theta)}{\kappa_2^2\beta_1},
\label{eq:IGa1_map}
\end{equation}
and
\begin{equation}
c_{\text{IGa1}}=\frac{\beta_1}{\beta e^{-\beta/\beta_1}(q\beta_1+\beta)\Gamma(q)}.
\end{equation}
Its CDF can be written in terms of upper incomplete gamma function $\Gamma(a,z)$ \cite{nist2026digital} as
\begin{equation}
F_{IGa1}(x;\beta,\beta_1,q)
=
C_{\text{IGa1}}\left[\beta\,\Gamma\!\left(q,\frac{\beta}{x}-\frac{\beta}{\beta_1}\right)
+\beta_1\Gamma\!\left(q+1,\frac{\beta}{x}-\frac{\beta}{\beta_1}\right)\right],
\label{eq:IGa1_cdf}
\end{equation}
with
\begin{equation}
C_{\text{IGa1}}=\frac{1}{(\beta+q\beta_1)\Gamma(q)}.
\end{equation}
Thus, the \(\kappa=0\) reduction keeps the finite upper endpoint while
isolating the IGa1 stochastic sector.

\subsubsection{Heavy-tail branch: the \texorpdfstring{$\kappa_1=0$}{kappa1=0} Beta-prime / Beta type-II limit}

Setting
\begin{equation}
\kappa_1=0
\qquad
\Longleftrightarrow
\qquad
\beta_1\to\infty
\end{equation}
removes the finite upper boundary. The master equation
\eqref{eq:master_sde_kappa} then becomes
\begin{equation}
dx
=
-\gamma(x-\theta)\,dt
+
\sqrt{\kappa^2 x+\kappa_2^2 x^2}\,dW_t
=
-\gamma(x-\theta)\,dt
+
\sqrt{\kappa^2 x\left(1+\frac{x}{\beta_2}\right)}\,dW_t .
\label{eq:mB2_sde}
\end{equation}
In the notation used in Table~\ref{table1}, the corresponding stationary
branch is denoted by BP/B2, with density \cite{dashti2021combined}
\begin{equation}
f_{BP/B2}(x;\beta_2,p,q)
=
c_{\text{BP}}\left(1+\frac{x}{\beta_2}\right)^{-p-q-1}
\left(\frac{x}{\beta_2}\right)^{p-1},
\qquad x\ge 0,
\label{eq:mB2_pdf}
\end{equation}
where
\begin{equation}
c_{\text{BP}}=\frac{1}{\beta_2 B(p,q+1)}.
\end{equation}
Its cumulative distribution function is
\begin{equation}
F_{BP/B2}(x;\beta_2,p,q)
=
I\!\left(\frac{x}{x+\beta_2};\,p,\,q+1\right),
\label{eq:mB2_cdf}
\end{equation}
with parameter map
\begin{equation}
p=\frac{2\gamma\theta}{\kappa^2},
\qquad
q=\frac{2\gamma}{\kappa_2^2}
=\frac{2\gamma\beta_2}{\kappa^2}.
\label{eq:mB2_map}
\end{equation}
This is the heavy-tail BP/B2 branch of the hierarchy.

\subsubsection{Compact-support branch: the \texorpdfstring{$\kappa_2=0$}{kappa2=0} Beta type-I limit}

Setting
\begin{equation}
\kappa_2=0
\qquad
\Longleftrightarrow
\qquad
\beta_2\to\infty
\end{equation}
removes the heavy-tail contribution. The master equation
\eqref{eq:master_sde_kappa} then reduces to
\begin{equation}
dx
=
-\gamma(x-\theta)\,dt
+
\sqrt{\kappa^2 x\left(1-\kappa_1^2 x\right)}\,dW_t
=
-\gamma(x-\theta)\,dt
+
\sqrt{\kappa^2 x\left(1-\frac{x}{\beta_1}\right)}\,dW_t .
\label{eq:B1_sde}
\end{equation}
In the notation of Table~\ref{table1}, the corresponding stationary density is
\begin{equation}
f_{B1}(x;\beta_1,p,q)
=
c_{\text{B1}}\left(\frac{x}{\beta_1}\right)^{p-1}
\left(1-\frac{x}{\beta_1}\right)^{q-1},
\qquad 0\le x\le \beta_1,
\label{eq:B1_pdf}
\end{equation}
with
\begin{equation}
c_{\text{B1}}=\frac{1}{\beta_1B(p,q)}.
\end{equation}
Its cumulative distribution function is
\begin{equation}
F_{B1}(x;\beta_1,p,q)=I\!\left(\frac{x}{\beta_1};p,q\right).
\label{eq:B1_cdf}
\end{equation}
The parameters remain
\begin{equation}
p=\frac{2\gamma\theta}{\kappa^2},
\qquad
q=\frac{2\gamma(1-\kappa_1^2\theta)}{\kappa^2\kappa_1^2}
=\frac{2\gamma(\beta_1-\theta)}{\kappa^2}.
\label{eq:B1_map}
\end{equation}
Thus, the \(\kappa_2=0\) reduction preserves compact support while removing
the heavy-tail stochastic sector.

\subsubsection{Interpretation of the one-parameter reductions}

In the present ordering, the one-parameter hierarchy reads
\begin{equation}
mB
\;\xrightarrow{\;\kappa=0\;}\;
IGa1,
\qquad
mB
\;\xrightarrow{\;\kappa_1=0\;}\;
BP/B2,
\qquad
mB
\;\xrightarrow{\;\kappa_2=0\;}\;
B1 .
\label{eq:one_param_hierarchy}
\end{equation}
The \(\kappa=0\) branch isolates the finite-support inverse-Gamma-type
sector, the \(\kappa_1=0\) branch produces the heavy-tail Beta-prime/Beta
 type-II sector, and the \(\kappa_2=0\) branch produces the compact-support
Beta type-I sector.

\subsection{Two-parameter limits: Gamma and inverse-Gamma endpoints}

The one-parameter reductions discussed above already separate the
finite-support inverse-Gamma branch, the heavy-tail branch, and the
compact-support Beta branch of the hierarchy. The next step is to follow
these branches to their simplest endpoints. We again begin with the
\(\kappa=0\) branch.

\subsubsection{Inverse-Gamma endpoint from the \texorpdfstring{$\kappa=0$}{kappa=0} branch}

Starting from the finite-support inverse-Gamma branch
\eqref{eq:IGa1_sde}, we remove the upper boundary by setting
\begin{equation}
\kappa_1=0
\qquad
\Longleftrightarrow
\qquad
\beta_1\to\infty .
\end{equation}
This yields the purely multiplicative process
\begin{equation}
dx
=
-\gamma(x-\theta)\,dt
+
\kappa_2 x\,dW_t .
\label{eq:IGa_sde}
\end{equation}
In the notation of Table~\ref{table1}, the stationary inverse-Gamma law is
\begin{equation}
f_{IGa}(x;\beta,q)
=
c_{\text{IGa}}\left(\frac{x}{\beta}\right)^{-q-2}e^{-\beta/x},
\qquad x>0,
\label{eq:IGa_pdf}
\end{equation}
with
\begin{equation}
c_{\text{IGa}}=\frac{1}{\beta\Gamma(q+1)},
\end{equation}
and
\begin{equation}
F_{IGa}(x;\beta,q)=C_{\text{IGa}}\Gamma\!\left(q+1,\frac{\beta}{x}\right),
\label{eq:IGa_cdf}
\end{equation}
with
\begin{equation}
C_{\text{IGa}}=\frac{1}{\Gamma(q+1)}.
\end{equation}
This is the power-law-tailed terminal endpoint of the multiplicative-noise
sector:
\begin{equation}
IGa1
\;\xrightarrow{\;\kappa_1=0\;}
IGa .
\label{eq:IGa_IGa1_endpoint}
\end{equation}

\subsubsection{Terminal limits of the heavy-tail branch}

Starting from the heavy-tail branch \eqref{eq:mB2_sde}, there are two distinct
terminal reductions. First, if one sets 
\begin{equation}
\kappa=0,
\end{equation}
one again arrives at (\ref{eq:IGa_sde}) and, as a result, reaches the inverse-Gamma endpoint already discussed above:
\begin{equation}
BP/B2
\;\xrightarrow{\;\kappa=0\;}\;
IGa .
\label{eq:IGa_B2_endpoint}
\end{equation}

If instead one sets
\begin{equation}
\kappa_2=0,
\end{equation}
the quadratic stochastic contribution is removed and one obtains
\begin{equation}
dx
=
-\gamma(x-\theta)\,dt
+
\kappa \sqrt{x}\,dW_t .
\label{eq:Ga_sde_from_mB2}
\end{equation}
The corresponding exponentially suppressed endpoint is denoted
Ga in Table~\ref{table1}, with stationary density
\begin{equation}
f_{\text{Ga}}(x;p,\lambda)
=
c_{\text{Ga}}\left(\frac{x}{\lambda}\right)^{p-1}e^{-x/\lambda},
\qquad x>0,
\label{eq:Ga_pdf_from_mB2}
\end{equation}
where
\begin{equation}
c_{\text{Ga}}=\frac{1}{\lambda \Gamma(p)},
\qquad
\lambda=\frac{\theta}{p}=\frac{\kappa^2}{2\gamma},
\label{eq:Ga_map_from_mB2}
\end{equation}
The corresponding CDF is given, in terms of lower incomplete gamma function $\gamma(a,z)$ \cite{nist2026digital}, by
\begin{equation}
F_{\text{Ga}}(x;p,\lambda)
=
C_{\text{Ga}}\gamma\!\left(p,\frac{x}{\lambda}\right),
\qquad
C_{\text{Ga}}=\frac{1}{\Gamma(p)}.
\label{eq:Ga_cdf_from_mB2}
\end{equation}
Thus the heavy-tail branch has two terminal endpoints,
\begin{equation}
BP/B2
\;\xrightarrow{\;\kappa=0\;}\;
IGa,
\qquad
BP/B2
\;\xrightarrow{\;\kappa_2=0\;}\;
Ga .
\label{eq:mB2_two_terminal_limits}
\end{equation}
An important clarification to (\ref{eq:IGa_IGa1_endpoint}) and (\ref{eq:mB2_two_terminal_limits}) must be made. In the former case the reduction can be made at the PDF level by simply setting $\beta_1\to \infty$, which is equivalent to $\kappa_1=0$, in (\ref{eq:IGa1_pdf}). In the latter case, however, the reduction must occur at the SDE level: it is setting $\kappa=0$ and $\kappa_1=0$ in (\ref{eq:master_sde_kappa}) that produces (\ref{eq:IGa_sde}) and (\ref{eq:IGa_pdf}) and it is setting $\kappa_2=0$ and $\kappa_1=0$ that produces (\ref{eq:Ga_sde_from_mB2}) and (\ref{eq:Ga_pdf_from_mB2}) -- not literally setting $\kappa=0$ or $\kappa_2=0$ in (\ref{eq:mB2_pdf}). On the other hand, (\ref{eq:mB2_pdf}) yields $f_{BP/B2} \propto (x/\beta_2)^{-q-2}, x\gg\beta_2$ and $f_{BP/B2} \propto (x/\beta_2)^{p-1}, x\ll\beta_2$. For IGa endpoint the former yields power-law tails in (\ref{eq:IGa_pdf}), while the latter mimics exponential decay as $p \to \infty$ when $\kappa=0$ per (\ref{eq:mB2_map}); for Ga endpoint the former mimics exponential decay in (\ref{eq:Ga_pdf_from_mB2}), as $q \to \infty$ when $\kappa_2=0$ per (\ref{eq:mB2_map}), while latter gives the initial power-law dependence.

\subsubsection{Ga endpoint from the compact-support branch}

Starting from the compact-support branch \eqref{eq:B1_sde}, we remove the
upper boundary by setting
\begin{equation}
\kappa_1=0
\qquad
\Longleftrightarrow
\qquad
\beta_1\to\infty .
\end{equation}
This again yields (\ref{eq:Ga_sde_from_mB2}) and 
consequently the same exponentially suppressed endpoint, whose stationary  PDF and CDF are  given by Eqs.~\eqref{eq:Ga_pdf_from_mB2}--\eqref{eq:Ga_cdf_from_mB2}, is reached from the compact-support branch:
\begin{equation}
B1
\;\xrightarrow{\;\kappa_1=0\;}\;
Ga .
\label{eq:B1_to_Ga}
\end{equation}
Insofar as Ga reduction directly from \eqref{eq:B1_pdf}, we notice that for $\beta_1 \to \infty$ we have from (\ref{eq:B1_map}), $q-1 \to 2 \gamma \beta_1 / \kappa^2$ so that $(1-\frac{x}{\beta_1})^{q-1} \to e ^{-2\gamma x/ \kappa^2}$.

\subsubsection{Summary of the terminal structure}

Collecting the results above, the Beta-sector hierarchy is
summarized schematically as
\begin{equation}
mB
\;\xrightarrow{\;\kappa=0\;}\;
IGa1
\;\xrightarrow{\;\kappa_1=0\;}\;
IGa ,
\label{eq:mB_IGa1_IGa}
\end{equation}
\begin{equation}
mB
\;\xrightarrow{\;\kappa_1=0\;}\;
BP/B2
\;\xrightarrow{\;\kappa=0\;}\;
IGa ,
\qquad
mB
\;\xrightarrow{\;\kappa_1=0\;}\;
BP/B2
\;\xrightarrow{\;\kappa_2=0\;}\;
Ga ,
\label{eq:mB_mB2_limits}
\end{equation}
and
\begin{equation}
mB
\;\xrightarrow{\;\kappa_2=0\;}\;
B1
\;\xrightarrow{\;\kappa_1=0\;}\;
Ga .
\label{eq:mB_B1_Ga}
\end{equation}
IGa is therefore the terminal power-law-tailed
endpoint of the multiplicative-noise sector, while Ga is the common
exponentially suppressed endpoint of both the compact-support and heavy-tail
branches. In this notation, IGa1 is the finite-support precursor of
IGa, just as B1 and BP/B2 both flow to Ga under further
reductions.

\subsection{From the Beta hierarchy to the generalized hierarchy}

The previous subsections established the hierarchy generated by the master stochastic differential equation in the original Beta sector. We now extend that hierarchy to a generalized one by introducing the power transformation
\begin{equation}
y=x^\alpha,
\qquad
\alpha>0 .
\label{eq:power_transform}
\end{equation}
This is the step that connects the mB family to generalized extensions and, in limiting cases, to the familiar generalized Beta type-I (GB1), generalized Beta Prime (GBP/GB2), generalized Gamma (GGa), and GIGa families.

If $f_Y(y)$ is a density in the Beta sector, then
\begin{equation}
f_Y(y)\,dy=f_X(x)\,dx,
\end{equation}
so the transformed density is
\begin{equation}
f_X(x)=\alpha x^{\alpha-1} f_Y(x^\alpha).
\label{eq:density_pushforward}
\end{equation}
Thus every Beta-sector branch generates a corresponding generalized branch under the map $y=x^\alpha$.

In the generalized sections below we keep $\beta_1$ and $\beta_2$ as the scale parameters inherited from the parent Beta-sector variable $y$; we do \emph{not} relabel $\beta_i^{1/\alpha}\to\beta_i$. Accordingly, a bounded branch on $0\le y\le \beta_1$ is mapped to
\begin{equation}
0\le x\le \beta_1^{1/\alpha},
\label{eq:generalized_support_convention}
\end{equation}
whereas an unbounded branch on $y\ge 0$ remains supported on $x\ge 0$. With this convention the generalized formulas are written naturally in terms of $x^\alpha/\beta_i$. Since $y=x^\alpha$, the combinations $y/\beta_i$ appearing in the parent distribution become $x^\alpha/\beta_i$ after the transformation. Throughout the following derivation, $\beta_i$ retain their original definitions and are not redefined as $\beta_i^{1/\alpha}$.

What changes under the transformation is the local and asymptotic scaling. For example,
\begin{equation}
f_Y(y)\sim y^{p-1}
\quad (y\to 0)
\qquad \Longrightarrow \qquad
f_X(x)\sim x^{\alpha p-1}
\quad (x\to 0),
\label{eq:origin_scaling_alpha}
\end{equation}
and
\begin{equation}
f_Y(y)\sim y^{-r}
\quad (y\to\infty)
\qquad \Longrightarrow \qquad
f_X(x)\sim x^{-\alpha r+\alpha-1}
\quad (x\to\infty).
\label{eq:tail_scaling_alpha}
\end{equation}
Hence the parameter $\alpha$ reshapes endpoint and tail exponents while preserving the overall hierarchy.

There are two inequivalent ways to implement the transformation \eqref{eq:power_transform}.

\paragraph{(i) Dynamical route}
One may apply the transformation at the level of the stochastic differential equation itself. That is, one treats $y=x^\alpha$ as the variable obeying the Beta-sector dynamics and rewrites the result as an SDE for $x$. The stationary solutions of that transformed SDE define the dynamical generalized hierarchy. Because nonlinear changes of variable act nontrivially on stochastic equations, this route is not, in general, equivalent to transforming the stationary distributions afterward.

\paragraph{(ii) Distributional route}
Alternatively, one may first derive the stationary distributions in the Beta sector and then push them forward under the transformation $y=x^\alpha$. This produces a second generalized hierarchy obtained directly from the modified-Beta stationary family.

The distinction between these two constructions is one of the central points of this work. In many limiting cases they reproduce the same standard generalized families, but at the top level they lead to different generalized extensions. This allows us to separate properties intrinsic to the hierarchy from those that depend on whether the generalization is performed before or after stationary-state reduction. In the appendix summary, Table~\ref{table2} records the dynamical route, whereas Table~\ref{table3} is obtained by applying the pushforward rule \eqref{eq:density_pushforward} branch by branch to the Beta-sector summary in Table~\ref{table1}.

We begin with the dynamical route, since it stays closest to the stochastic origin of the hierarchy: the power transformation is implemented directly at the level of the differential equation, and the generalized steady states are then derived from the transformed dynamics. We then turn to the distributional route, where the already known stationary modified-Beta family is pushed forward under the same transformation.

\section{Generalized hierarchy from the transformed stochastic differential equation (Dynamical route)}

We now construct the generalized hierarchy by implementing the power
transformation directly at the level of the stochastic differential equation.
This route is the natural continuation of the master-equation construction
developed in the Beta sector: the dynamics are transformed first, and the
generalized steady states are then derived from the resulting equation.

To implement the dynamical route, we regard
\begin{equation}
y=x^\alpha,
\qquad
\alpha>0 ,
\label{eq:y_equals_xalpha_dyn}
\end{equation}
as the stochastic variable obeying the Beta-sector dynamics. After rewriting
the resulting process in terms of \(x\), one obtains a generalized
stochastic differential equation. In the original
\((\kappa,\kappa_1,\kappa_2)\) parametrization, with $\gamma$, $\kappa_1$, and $\kappa_2$ renormalized by $1/\alpha$,
 the \emph{formally} transformed dynamics (see \ref{Ito}) can
be written as
\begin{equation}
dx
=
-\gamma\bigl(x-\theta x^{1-\alpha}\bigr)\,dt
+
\sqrt{
\left(1-\kappa_1^2 x^\alpha\right)
\left(\kappa^2 x^{2-\alpha}+\kappa_2^2 x^2\right)
}\,dW_t ,
\label{eq:mGB_sde_dyn_kappa}
\end{equation}
and in the rescaled form of the theory, this becomes
\begin{equation}
dx
=
-\gamma\bigl(x-\theta x^{1-\alpha}\bigr)\,dt
+
\sqrt{\kappa^2
x^{2-\alpha}
\left(
1-\frac{x^\alpha}{\beta_1}
\right)
\left(
1+\frac{x^\alpha}{\beta_2}
\right)
}\,dW_t .
\label{eq:mGB_sde_dyn}
\end{equation}
This is the direct generalized counterpart of the master modified-Beta
dynamics. In what follows, we use the notation and exponent structure summarized in Table~\ref{table2}; in particular, the main-text formulas are written to match the $q_1$-based dynamical hierarchy recorded there. To keep the notation parallel with the Beta sector and the distributional route, we write the dynamical densities and cumulative distribution functions as $f$ and $F$ rather than introducing a separate notation.

\subsection{Top-level distribution from the dynamical route: the \texorpdfstring{$mGB$}{mGB} family}

To write the top-level dynamical family compactly, it is convenient to supplement the Beta-sector parameters $p$ and $q$ of Eq.~\eqref{eq:pq_map} with the combination
\begin{equation}
q_1\equiv p+q
=
\frac{2\gamma(\kappa^2+\theta\kappa_2^2)}{\kappa^2(\kappa^2\kappa_1^2+\kappa_2^2)} .
\label{eq:q1_def_dyn}
\end{equation}
The equality $q_1=p+q$ makes transparent that the generalized dynamical exponents reduce to the Beta-sector ones when $\alpha=1$.

The steady-state density associated with Eq.~\eqref{eq:mGB_sde_dyn} defines the top-level member of mGB family:
\begin{equation}
f_{mGB}(x;\alpha,\beta_1,\beta_2,p,q,q_1)
=
c_{\text{mGB}}\,\left(\frac{x^{\alpha}}{\beta_2}\right)^{\frac{p+\alpha-2}{\alpha}} 
\left(1-\frac{x^{\alpha}}{\beta_1}\right)^{\frac{q}{\alpha}-1}
\left(1+\frac{x^{\alpha}}{\beta_2}\right)^{-\frac{q_1}{\alpha}-1},
\qquad 0\le x\le \beta_1^{1/\alpha} ,
\label{eq:mGB_pdf_dyn}
\end{equation}
where $p+\alpha>1$ is required to ensure integrability at the lower endpoint. This is the same top-level mGB structure, with the normalization constant $c_{\text{mGB}}$, which is summarized in Table~\ref{table2}: the role of $q_1$ is to encode the heavy-tail crossover exponent of the dynamical route, while the left-end and finite-support exponents are shifted by the transformed SDE itself. The corresponding CDF can be written in terms of Appell function $F_1$ \cite{nist2026digital} as
\begin{equation}
F_{mGB}(x;\alpha,\beta_1,\beta_2,p,q,q_1)
=
C_{\text{mGB}}\,\left(\frac{x^{\alpha}}{\beta_2}\right)^{\frac{p+\alpha-1}{\alpha}} 
F_1\!\left(
\frac{p+\alpha-1}{\alpha};
1-\frac{q}{\alpha},
1+\frac{q_1}{\alpha};
1+\frac{p+\alpha-1}{\alpha};
\frac{x^{\alpha}}{\beta_1},
-\frac{x^{\alpha}}{\beta_2}
\right),
\label{eq:mGB_cdf_dyn}
\end{equation}
with
\begin{equation}
C_{\text{mGB}}=\frac{c_{\text{mGB}}\,\beta_2^{1/\alpha}}{p+\alpha-1} .
\label{eq:CmGB_def}
\end{equation}

\subsection{Interpretation}

The dynamical route shows that the generalized hierarchy can be obtained directly as the steady-state hierarchy of a transformed stochastic differential equation. Its top-level density exhibits the characteristic left-end exponent $p+\alpha-2$, the finite-support exponent $q/\alpha-1$ (exponent of $\left(\beta^{1/\alpha}-x\right)$ on approach to $\beta^{1/\alpha}$), and the crossover exponent $-q_1/\alpha-1$ ($(q_1-\alpha)$ being the right-hand exponent of heavy tail), all arising from the transformed dynamics itself. Since $q_1=p+q$, the $\alpha\to 1$ limit returns the Beta-sector exponents $(p-1,q-1,-p-q-1)$ as it should. The significance of this structure becomes clearer when it is compared with the hierarchy obtained by power-transforming the stationary modified-Beta family. We return to that comparison after constructing the distributional route in the next section.

\subsection{One-parameter reductions of the dynamical hierarchy}

To parallel the organization of the Beta hierarchy, we present the lower
branches of the dynamical generalized hierarchy in the order
\(\kappa=0\), \(\kappa_1=0\), and \(\kappa_2=0\).

\subsubsection{Generalized finite-support inverse-Gamma branch: the \texorpdfstring{$\kappa=0$}{kappa=0} reduction}

Setting
\begin{equation}
\kappa=0
\end{equation}
in Eq.~\eqref{eq:mGB_sde_dyn_kappa} removes the square-root stochastic
contribution inherited from the linear-in-\(y\) noise and leaves
\begin{equation}
dx
=
-\gamma\bigl(x-\theta x^{1-\alpha}\bigr)\,dt
+
\kappa_2 x
\sqrt{1-\kappa_1^2 x^\alpha}\,dW_t
=
-\gamma\bigl(x-\theta x^{1-\alpha}\bigr)\,dt
+
\kappa_2 x
\sqrt{1-\frac{x^\alpha}{\beta_1}}\,dW_t .
\label{eq:GIGa1_sde_dyn}
\end{equation}
This is the generalized finite-support inverse-Gamma branch of the dynamical route,
denoted by GIGa1 in Table~\ref{table2}. Its normalized PDF is collected in the summary
table. The CDF can be written as the corresponding integral of the PDF, but no
compact closed-form expression is listed here.

\subsubsection{Generalized heavy-tail branch: the \texorpdfstring{$\kappa_1=0$}{kappa1=0} reduction}

Setting
\begin{equation}
\kappa_1=0
\qquad
\Longleftrightarrow
\qquad
\beta_1\to\infty
\end{equation}
removes the finite upper boundary. In the notation of Table~\ref{table2},
this branch is denoted by GBP. Using the shorthand $q_1$ introduced in
Eq.~\eqref{eq:q1_def_dyn}, its stationary density is
\begin{equation}
f_{GBP}(x;\alpha,\beta_2,p,q_1)
=
c_{\text{GBP}}\left(\frac{x^{\alpha}}{\beta_2}\right)^{\frac{p+\alpha-2}{\alpha}} 
\left(1+\frac{x^{\alpha}}{\beta_2}\right)^{-\frac{q_1}{\alpha}-1},
\qquad x\ge 0,
\label{eq:mGB2_pdf_dyn}
\end{equation}
with CDF
\begin{equation}
F_{GBP}(x;\alpha,\beta_2,p,q_1)
=
C_{\text{GBP}}\left(\frac{x^{\alpha}}{\beta_2}\right)^{\frac{p+\alpha-1}{\alpha}}{}_2F_1\!\left(
1+\frac{q_1}{\alpha},\,
\frac{p+\alpha-1}{\alpha},\,
1+\frac{p+\alpha-1}{\alpha},\,
-\frac{x^{\alpha}}{\beta_2}
\right),
\label{eq:mGB2_cdf_dyn}
\end{equation}
where constants $c_{\text{GBP}}$ and $C_{\text{GBP}}$ are those given
in Table~\ref{table2} and ${}_2F_1$ is the hypergeometric function \cite{nist2026digital}.

\subsubsection{Generalized compact-support branch: the \texorpdfstring{$\kappa_2=0$}{kappa2=0} reduction}

Setting
\begin{equation}
\kappa_2=0
\qquad
\Longleftrightarrow
\qquad
\beta_2\to\infty
\end{equation}
removes the heavy-tail crossover factor. In the notation of
Table~\ref{table2}, the resulting compact-support branch is denoted by
GB1, with stationary density
\begin{equation}
f_{GB1}(x;\alpha,\beta_1,p,q)
=
c_{\text{GB1}}\left(\frac{x^{\alpha}}{\beta_1}\right)^{\frac{p+\alpha-2}{\alpha}}\left(1-\frac{x^{\alpha}}{\beta_1}\right)^{\frac{q}{\alpha}-1},
\qquad 0\le x\le \beta_1^{1/\alpha},
\label{eq:GB1_pdf_dyn}
\end{equation}
and CDF
\begin{equation}
F_{GB1}(x;\alpha,\beta_1,p,q)
=
C_{\text{GB1}}\left(\frac{x^{\alpha}}{\beta_1}\right)^{\frac{p+\alpha-1}{\alpha}}{}_2F_1\!\left(
1-\frac{q}{\alpha},\,
\frac{p+\alpha-1}{\alpha},\,
1+\frac{p+\alpha-1}{\alpha},\,
\frac{x^{\alpha}}{\beta_1}
\right).
\label{eq:GB1_cdf_dyn}
\end{equation}
This branch retains the exponent shift characteristic of the dynamical route.

\subsubsection{Interpretation of the one-parameter reductions}

In the present ordering, the dynamical generalized hierarchy reads
\begin{equation}
mGB
\;\xrightarrow{\;\kappa=0\;}\;
GIGa1,
\qquad
mGB
\;\xrightarrow{\;\kappa_1=0\;}\;
GBP,
\qquad
mGB
\;\xrightarrow{\;\kappa_2=0\;}\;
GB1 .
\label{eq:mGB_dyn_one_param_hierarchy}
\end{equation}

\subsection{Two-parameter limits}

The terminal limits of the dynamical hierarchy follow the same tree structure as before. We therefore begin with the $\kappa=0$ branch and then discuss the terminal reductions of the heavy-tail and compact-support branches.

\subsubsection{Generalized inverse-Gamma endpoint from the \texorpdfstring{$\kappa=0$}{kappa=0} branch}

Starting from the generalized finite-support inverse-Gamma branch \eqref{eq:GIGa1_sde_dyn}, we remove the upper boundary by setting
\begin{equation}
\kappa_1=0
\qquad
\Longleftrightarrow
\qquad
\beta_1\to\infty .
\end{equation}
This yields
\begin{equation}
dx
=
-\gamma\bigl(x-\theta x^{1-\alpha}\bigr)\,dt
+
\kappa_2 x\,dW_t ,
\label{eq:GIGa_sde_dyn}
\end{equation}
whose stationary distribution is the generalized inverse-Gamma law denoted by GIGa in Table~\ref{table2}:
\begin{equation}
f_{GIGa}(x;\alpha,\beta,\lambda_1)
=
c_{\text{GIGa}}\left(\frac{\alpha x^{\alpha}}{\beta}\right)^{-\frac{1/\lambda_1+2}{\alpha}}e^{-\frac{\beta}{\alpha x^{\alpha}}},
\qquad x>0,
\label{eq:GIGa_pdf_dyn}
\end{equation}
with CDF
\begin{equation}
F_{GIGa}(x;\alpha,\beta,\lambda_1)
=
C_{\text{GIGa}}\Gamma\left(\frac{1+1/\lambda_1}{\alpha},\frac{\beta}{\alpha x^{\alpha}}\right),
\label{eq:GIGa_cdf_dyn}
\end{equation}
where $\lambda_1=\theta/\beta$ and constants are listed in Table~\ref{table2}. This is the terminal power-law-tailed endpoint of the generalized multiplicative-noise sector:
\begin{equation}
GIGa1
\;\xrightarrow{\;\kappa_1=0\;}\;
GIGa .
\end{equation}

\subsubsection{Terminal limits of the generalized heavy-tail branch}

Starting from the generalized heavy-tail branch, there are two distinct terminal reductions. If one sets
\begin{equation}
\kappa=0,
\end{equation}
one reaches the generalized inverse-Gamma endpoint already discussed above:
\begin{equation}
GBP
\;\xrightarrow{\;\kappa=0\;}\;
GIGa .
\end{equation}
If instead one sets
\begin{equation}
\kappa_2=0,
\end{equation}
one obtains the generalized Gamma endpoint denoted by GGa in Table~\ref{table2}, with stationary density
\begin{equation}
f_{GGa}(x;\alpha,p,\lambda_2)
=
c_{\text{GGa}}
\left(\frac{x^{\alpha}}{\alpha\lambda_2}\right)^{\frac{p+\alpha-2}{\alpha}}
e^{-\frac{x^{\alpha}}{\alpha\lambda_2}},
\qquad x\ge 0,
\label{eq:GGa_pdf_dyn}
\end{equation}
and CDF
\begin{equation}
F_{GGa}(x;\alpha,p,\lambda_2)
=
1-C_{\text{GGa}}\Gamma\left( \frac{p+\alpha -1}{\alpha}, \frac{x^{\alpha}}{\alpha \lambda_2}\right),
\label{eq:GGa_cdf_dyn}
\end{equation}
where $\lambda_2=\theta/p$. Thus,
\begin{equation}
GBP
\;\xrightarrow{\;\kappa=0\;}\;
GIGa,
\qquad
GBP
\;\xrightarrow{\;\kappa_2=0\;}\;
GGa .
\label{eq:mGB2_two_terminal_limits_dyn}
\end{equation}

\subsubsection{Generalized Gamma endpoint from the compact-support branch}

Starting from the generalized compact-support branch \eqref{eq:GB1_pdf_dyn}, we remove the upper boundary by setting
\begin{equation}
\kappa_1=0
\qquad
\Longleftrightarrow
\qquad
\beta_1\to\infty ,
\end{equation}
which results in the same GGa endpoint \eqref{eq:GGa_pdf_dyn},
\begin{equation}
GB1
\;\xrightarrow{\;\kappa_1=0\;}\;
GGa .
\label{eq:GB1_to_GGa_dyn}
\end{equation}
Thus the generalized Gamma distribution is again the common exponentially suppressed endpoint of both the heavy-tail and compact-support branches.

\subsubsection{Summary of the generalized terminal structure}

Collecting the results above, the dynamical route produces the generalized hierarchy
\begin{equation}
mGB
\;\xrightarrow{\;\kappa=0\;}\;
GIGa1
\;\xrightarrow{\;\kappa_1=0\;}\;
GIGa ,
\label{eq:mGB_GIGa1_GIGa_dyn}
\end{equation}
\begin{equation}
mGB
\;\xrightarrow{\;\kappa_1=0\;}\;
GBP
\;\xrightarrow{\;\kappa=0\;}\;
GIGa ,
\qquad
mGB
\;\xrightarrow{\;\kappa_1=0\;}\;
GBP
\;\xrightarrow{\;\kappa_2=0\;}\;
GGa ,
\label{eq:mGB_mGB2_limits_dyn}
\end{equation}
and
\begin{equation}
mGB
\;\xrightarrow{\;\kappa_2=0\;}\;
GB1
\;\xrightarrow{\;\kappa_1=0\;}\;
GGa .
\label{eq:mGB_GB1_GGa_dyn}
\end{equation}

\subsection{Interpretation}

The dynamical route shows that the generalized hierarchy can arise directly as the steady-state hierarchy of a transformed stochastic equation, not only as a power deformation of already known stationary distributions. In this sense, it places the generalized families on the same conceptual footing as the original Beta hierarchy and makes the stochastic role of the deformation parameter $\alpha$ fully explicit.

Having established this route, we now turn to the complementary construction. Instead of transforming the stochastic dynamics first, we keep the Beta-sector dynamics fixed and apply the same power transformation directly to the stationary modified-Beta family. This leads to the second generalized hierarchy, denoted $mB_{\alpha}$.

\section{Generalized hierarchy from the power transformation of the modified-Beta family (Distributional route)}

We now turn to the second route to the generalized hierarchy, obtained by directly power-transforming the stationary distributions of the Beta sector. Here the stochastic dynamics in the original sector are taken as already solved, and the generalized family is constructed by pushing forward the modified-Beta hierarchy under the transformation
\begin{equation}
y=x^\alpha,
\qquad
\alpha>0 .
\label{eq:y_xalpha_again}
\end{equation}
In the notation of Table~\ref{table3}, the resulting top-level family is denoted by $\text{mB}_{\alpha}$. Throughout this section, Table~\ref{table3} should be read as the direct image of Table~\ref{table1} under the pushforward rule \eqref{eq:density_pushforward}; unlike Table~\ref{table2}, it is not obtained from a separately transformed stochastic equation.

\subsection{Top-level distribution: the \texorpdfstring{$mB_{\alpha}$}{mBalpha} family}

Starting from the modified-Beta density in the variable $y$, the corresponding density in the variable $x$ is obtained from
\begin{equation}
f_{mB}(y)\,dy=f_{mB_{\alpha}}(x)\,dx,
\qquad
y=x^\alpha,
\qquad
dy=\alpha x^{\alpha-1}\,dx .
\end{equation}
The top-level density of the power-transformed hierarchy is therefore
\begin{equation}
f_{mB_{\alpha}}(x;\alpha,\beta_1,\beta_2,p,q)
=
c_{\text{mB}}\left(\frac{x^{\alpha}}{\beta_1}\right)^{\frac{\alpha p-1}{\alpha}}
\left(1-\frac{x^{\alpha}}{\beta_1}\right)^{q-1}
\left(1+\frac{x^{\alpha}}{\beta_2}\right)^{-p-q-1},
\qquad 0\le x\le \beta_1^{1/\alpha},
\label{eq:mBalpha_pdf_dist}
\end{equation}
with the normalization constant $c_{\text{mB}}$ given in Table~\ref{table3} \footnote{As in Beta Sector (\ref{eq:mB_pdf}), it may be more appropriate to scale left end of the distribution (first parenthesis in (\ref{eq:mBalpha_pdf_dist})) with $\beta_2$.}. Equivalently, this row of Table~\ref{table3} is obtained by applying \eqref{eq:density_pushforward} directly to the top row of Table~\ref{table1}. Its CDF is the direct power-transformed image of the Beta-sector CDF:
\begin{align}
F_{mB_{\alpha}}(x;\alpha,\beta_1,\beta_2,p,q)
&=
I\!\left(\frac{x^{\alpha}/\beta_1+x^{\alpha}/\beta_2}{1+x^{\alpha}/\beta_2};p,q\right)
\nonumber\\[4pt]
&\quad
+\frac{1}{B(p,q)\left[q+\dfrac{\beta_2}{\beta_1}(p+q)\right]}
\left(\frac{1-x^{\alpha}/\beta_1}{1+x^{\alpha}/\beta_2}\right)^q
\left(\frac{(1+\beta_1/\beta_2)x^{\alpha}/\beta_1}{1+x^{\alpha}/\beta_2}\right)^p .
\label{eq:mBalpha_cdf_dist}
\end{align}

\subsection{Comparison with the dynamical route}

The distributional-route hierarchy should now be compared with the generalized hierarchy obtained directly from the transformed stochastic differential equation. Because Table~\ref{table3} is fixed by the pushforward of Table~\ref{table1}, while Table~\ref{table2} is the independent stationary hierarchy generated by the transformed SDE, comparing the two isolates exactly the noncommutativity discussed above. The two constructions share the same reduction graph and connect analogous branches, but this should not be interpreted as equality of the corresponding
distributions. For \(\alpha\neq 1\), the dynamical and distributional routes
generally give different endpoint exponents and different normalization constants.
The top-level difference is the most visible manifestation of this route
dependence. \footnote{It should be emphasized that this noncommutativity and the resulting distinct hierarchies of distributions is due to the fact that Eq.~\eqref{eq:mGB_sde_dyn_kappa} was obtained by omitting the It\^o drift term due to its non-physicality. In \ref{Ito} it is shown that the addition of such term produces an SDE equation corresponding to the pushforward hierarchy.} 

The clearest differences appear in the three characteristic exponents. In the distributional route one finds
\begin{equation}
f_{mB_{\alpha}}(x)
\propto
\left(\frac{x^{\alpha}}{\beta_2}\right)^{\frac{\alpha p-1}{\alpha}}
\left(1-\frac{x^{\alpha}}{\beta_1}\right)^{q-1}
\left(1+\frac{x^{\alpha}}{\beta_2}\right)^{-p-q-1},
\label{eq:mBalpha}
\end{equation}
whereas in the dynamical route one obtains
\begin{equation}
f_{mGB}(x)
\propto
\left(\frac{x^{\alpha}}{\beta_2}\right)^{\frac{p+\alpha -2}{\alpha}}
\left(1-\frac{x^{\alpha}}{\beta_1}\right)^{\frac{q}{\alpha}-1}
\left(1+\frac{x^{\alpha}}{\beta_2}\right)^{-\frac{p+q}{\alpha}-1} .
\label{eq:mGB}
\end{equation}
Thus, while obviously converging to the same mB for $\alpha=1$, the two routes generally differ already in the left-end exponent, in the exponent governing the approach to the finite upper boundary, and in the heavy-tail crossover exponent. This exponent mismatch is the clearest local distinction between the dynamical and distributional routes.

\subsection{One-parameter reductions of the power-transformed hierarchy}

To remain parallel with the Beta hierarchy, we present the three one-parameter reductions in the order $\kappa=0$, $\kappa_1=0$, and $\kappa_2=0$. Each branch below is obtained by applying the pushforward rule \eqref{eq:density_pushforward} to the corresponding Beta-sector branch in Table~\ref{table1}.

\subsubsection{\texorpdfstring{$IGa1_{\alpha}$}{IGa1alpha}: the power-transformed \texorpdfstring{$\kappa=0$}{kappa=0} branch}

Starting from the finite-support inverse-Gamma branch IGa1 in the Beta sector and applying the power transformation yields
\begin{equation}
f_{IGa1_{\alpha}}(x;\alpha,\beta,\beta_1,q)
=
c_{\text{IGa1}}e^{-\beta/x^{\alpha}}\left(\frac{x^{\alpha}}{\beta}\right)^{\frac{-\alpha q-\alpha-1}{\alpha}}
\left(1-\frac{x^{\alpha}}{\beta_1}\right)^{q-1},
\qquad 0<x<\beta_1^{1/\alpha},
\label{eq:GIGa1_pdf_dist}
\end{equation}
with CDF
\begin{equation}
F_{IGa1_{\alpha}}(x;\alpha,\beta,\beta_1,q)
=
C_{\text{IGa1}}\left[\beta\,\Gamma\!\left(q,\frac{\beta}{x^{\alpha}}-\frac{\beta}{\beta_1}\right)
+\beta_1\Gamma\!\left(q+1,\frac{\beta}{x^{\alpha}}-\frac{\beta}{\beta_1}\right)\right].
\label{eq:GIGa1_cdf_dist}
\end{equation}
This is the finite-support inverse-Gamma branch, $\text{IGa}1_{\alpha}$, of the $\text{mB}_{\alpha}$ hierarchy.

\subsubsection{\texorpdfstring{$BP/B2_{\alpha}$}{BP/B2alpha}: the power-transformed \texorpdfstring{$\kappa_1=0$}{kappa1=0} branch}

Starting from the heavy-tail branch and then applying the power transformation yields
\begin{equation}
f_{BP/B2_{\alpha}}(x;\alpha,\beta_2,p,q)
=
c_{\text{BP}}\left(\frac{x^{\alpha}}{\beta_2}\right)^{\frac{\alpha p-1}{\alpha}}\left(1+\frac{x^\alpha}{\beta_2}\right)^{-p-q-1},
\qquad x\ge 0,
\label{eq:mGB2_pdf_dist}
\end{equation}
with CDF
\begin{equation}
F_{BP/B2_{\alpha}}(x;\alpha,\beta_2,p,q)
=
I\!\left(\frac{x^{\alpha}}{x^{\alpha}+\beta_2};\,p,\,q+1\right).
\label{eq:mGB2_cdf_dist}
\end{equation}
This is the power-transformed Beta-prime/Beta type-II branch.

\subsubsection{\texorpdfstring{$B1_{\alpha}$}{B1alpha}: the power-transformed \texorpdfstring{$\kappa_2=0$}{kappa2=0} branch}

Starting from the compact-support Beta type-I branch and power-transforming its stationary density yields
\begin{equation}
f_{B1_{\alpha}}(x;\alpha,\beta_1,p,q)
=
c_{\text{B1}}\left(\frac{x^{\alpha}}{\beta_1}\right)^{\frac{\alpha p-1}{\alpha}}\left(1-\frac{x^{\alpha}}{\beta_1}\right)^{q-1},
\qquad 0\le x\le \beta_1^{1/\alpha},
\label{eq:GB1_pdf}
\end{equation}
with CDF
\begin{equation}
F_{B1_{\alpha}}(x;\alpha,\beta_1,p,q)
=
I\!\left(\frac{x^{\alpha}}{\beta_1};p,q\right).
\label{eq:GB1_cdf}
\end{equation}
This is the compact-support $\text{B}1_{\alpha}$ branch listed in Table~\ref{table3}.

\subsubsection{Interpretation of the one-parameter reductions}

Thus the one-parameter power-transformed hierarchy takes the form
\begin{equation}
mB_{\alpha}
\;\xrightarrow{\;\kappa=0\;}\;
IGa1_{\alpha},
\qquad
mB_{\alpha}
\;\xrightarrow{\;\kappa_1=0\;}\;
BP/B2_{\alpha},
\qquad
mB_{\alpha}
\;\xrightarrow{\;\kappa_2=0\;}\;
B1_{\alpha} .
\label{eq:mGB_one_param_hierarchy}
\end{equation}

\subsection{Two-parameter limits}

The terminal limits of the power-transformed hierarchy follow the same tree structure as in the Beta sector, again with exponents reshaped by the Jacobian factor associated with $y=x^\alpha$.

\subsubsection{\texorpdfstring{$IGa_{\alpha}$}{IGaalpha} from the \texorpdfstring{$\kappa=0$}{kappa=0} branch}

Removing the upper boundary from $\text{IGa}1_{\alpha}$ gives the terminal inverse-Gamma-type endpoint
\begin{equation}
f_{IGa_{\alpha}}(x;\alpha,\beta,q)
=
c_{\text{IGa}}\left(\frac{x^{\alpha}}{\beta}\right)^{\frac{-\alpha q-\alpha-1}{\alpha}}e^{-\beta/x^{\alpha}},
\qquad x>0,
\label{eq:GIGa_pdf_dist}
\end{equation}
with CDF
\begin{equation}
F_{IGa_{\alpha}}(x;\alpha,\beta,q)
=
C_{\text{IGa}}\Gamma\!\left(q+1,\frac{\beta}{x^{\alpha}}\right).
\label{eq:GIGa_cdf_dist}
\end{equation}
Thus,
\begin{equation}
IGa1_{\alpha}
\;\xrightarrow{\;\kappa_1=0\;}\;
IGa_{\alpha} .
\end{equation}

\subsubsection{Terminal limits of the \texorpdfstring{$BP/B2_{\alpha}$}{BP/B2alpha} branch}

Starting from $\text{BP/B}2_{\alpha}$, there are again two terminal reductions. Setting $\kappa=0$ yields $\text{IGa}_{\alpha}$, while setting $\kappa_2=0$ yields the exponentially suppressed endpoint denoted by Ga$_{\alpha}$ in Table~\ref{table3}:
\begin{equation}
f_{\text{Ga}_{\alpha}}(x;\alpha,p,\lambda)
=
c_{\text{Ga}}\left(\frac{x^{\alpha}}{\lambda}\right)^{\frac{\alpha p-1}{\alpha}}e^{-x^{\alpha}/\lambda},
\qquad x\ge 0,
\label{eq:GGa_pdf_dist}
\end{equation}
with CDF
\begin{equation}
F_{\text{Ga}_{\alpha}}(x;\alpha,p,\lambda)
=
C_{\text{Ga}}\gamma\!\left(p,\frac{x^{\alpha}}{\lambda}\right).
\label{eq:GGa_cdf_dist}
\end{equation}
Thus,
\begin{equation}
BP/B2_{\alpha}
\;\xrightarrow{\;\kappa=0\;}\;
IGa_{\alpha},
\qquad
BP/B2_{\alpha}
\;\xrightarrow{\;\kappa_2=0\;}\;
\text{Ga}_{\alpha} .
\label{eq:mGB2_two_terminal_limits_dist}
\end{equation}

\subsubsection{\texorpdfstring{Ga$_{\alpha}$}{Gaalpha} from the compact-support branch}

Removing the upper boundary from B1$_{\alpha}$ produces the same terminal exponentially suppressed endpoint,
\begin{equation}
B1_{\alpha}
\;\xrightarrow{\;\kappa_1=0\;}\;
\text{Ga}_{\alpha} .
\label{eq:GB1_to_GGa_dist}
\end{equation}

\subsubsection{Summary of the generalized terminal structure}

Collecting the results above, the power-transformed route produces the hierarchy
\begin{equation}
mB_{\alpha}
\;\xrightarrow{\;\kappa=0\;}\;
IGa1_{\alpha}
\;\xrightarrow{\;\kappa_1=0\;}\;
IGa_{\alpha} ,
\label{eq:mGB_GIGa1_GIGa_dist}
\end{equation}
\begin{equation}
mB_{\alpha}
\;\xrightarrow{\;\kappa_1=0\;}\;
BP/B2_{\alpha}
\;\xrightarrow{\;\kappa=0\;}\;
IGa_{\alpha} ,
\qquad
mB_{\alpha}
\;\xrightarrow{\;\kappa_1=0\;}\;
BP/B2_{\alpha}
\;\xrightarrow{\;\kappa_2=0\;}\;
\text{Ga}_{\alpha} ,
\label{eq:mGB_mGB2_limits_dist}
\end{equation}
and
\begin{equation}
mB_{\alpha}
\;\xrightarrow{\;\kappa_2=0\;}\;
B1_{\alpha}
\;\xrightarrow{\;\kappa_1=0\;}\;
\text{Ga}_{\alpha} .
\label{eq:mGB_GB1_GGa_dist}
\end{equation}

\subsection{Interpretation}

The distributional route preserves the full tree structure of the Beta hierarchy while re-expressing it in the generalized notation of Table~\ref{table3}. In particular, the $\kappa=0$, $\kappa_1=0$, and $\kappa_2=0$ reductions remain visible after the power transformation, and the terminal endpoints IGa$_{\alpha}$ and Ga$_{\alpha}$ are reached in the same order as in the original hierarchy. Table~\ref{table3} is therefore best understood as the branch-by-branch pushforward of Table~\ref{table1}, not as an independent dynamical hierarchy. What changes is the Jacobian reshaping of the exponents, which is already visible in the top-level and lower-branch formulas.

With both generalized routes now in hand, the natural next step is to compare them with the historically standard Generalized Beta construction. This comparison clarifies which parts of the present framework recover the usual generalized-Beta landscape and which features are genuinely new consequences of the stochastic-hierarchical viewpoint.

\section{Relation to the traditional Generalized Beta family}

Having constructed both generalized routes, we now compare them with the traditional Generalized Beta (GB) family. This comparison clarifies both what is recovered from the standard generalized-Beta picture and what is genuinely new in the present stochastic construction.

\subsection{Traditional Generalized Beta distribution}

The traditional Generalized Beta distribution is introduced directly at the PDF level. To compare it with our notation, we keep the parent Beta-sector scales $\beta_1$ and $\beta_2$ rather than relabeling $\beta_i^{1/\alpha}\to\beta_i$. In this convention,
\begin{align}
f_{GB}(x;\alpha,\beta_1,\beta_2,p,q)
=
\frac{\alpha}{\beta_1^{1/\alpha} B(p,q)}
\left(1+\frac{\beta_1}{\beta_2}\right)^p
\left(\frac{x^{\alpha}}{\beta_1}\right)^{\frac{\alpha p-1}{\alpha}}
\left(1-\frac{x^{\alpha}}{\beta_1}\right)^{q-1}
\left(1+\frac{x^{\alpha}}{\beta_2}\right)^{-p-q}\nonumber \\
\quad 
=\frac{\alpha}{\beta_2^{1/\alpha} B(p,q)}
\left(1+\frac{\beta_2}{\beta_1}\right)^p
\left(\frac{x^{\alpha}}{\beta_2}\right)^{\frac{\alpha p-1}{\alpha}}
\left(1-\frac{x^{\alpha}}{\beta_1}\right)^{q-1}
\left(1+\frac{x^{\alpha}}{\beta_2}\right)^{-p-q},
\label{eq:GB_traditional}
\end{align}
with support
\begin{equation}
0\le x\le \beta_1^{1/\alpha} .
\end{equation}
Its CDF is
\begin{equation}
F_{GB}(x;\alpha,\beta_1,\beta_2,p,q)
=
I\!\left(
\frac{x^{\alpha}/\beta_1+x^{\alpha}/\beta_2}{1+x^{\alpha}/\beta_2};
\,p,\,q
\right),
\label{eq:GB_traditional_cdf}
\end{equation}
and the usual limits generate the standard hierarchy
\begin{equation}
GB
\;\xrightarrow{\;\beta_2\to\infty\;}\;
GB1,
\qquad
GB
\;\xrightarrow{\;\beta_1\to\infty\;}\;
GB2,
\label{eq:GB_main_branches}
\end{equation}
followed by the generalized Gamma and generalized inverse-Gamma branches. Thus the traditional GB family defines an algebraic hierarchy of distributions.

\subsection{What is new in the present framework}

The present construction starts not from a parametric PDF family, but from a master mean-reverting stochastic differential equation and its reductions. The hierarchy therefore appears as a hierarchy of stationary states. Its top-level Beta-sector distribution is the modified-Beta density mB, and the generalized extension is generated in two inequivalent ways: by applying the power transformation $y=x^\alpha$ directly at the SDE level before solving for the stationary state, or by applying the same transformation after obtaining the stationary distributions. In the notation of Tables~\ref{table2} and \ref{table3}, these two routes produce the dynamical-route mGB family and the distributional-route mB$_{\alpha}$ family, respectively. The important structural distinction is that Table~\ref{table2} is an independent stationary hierarchy generated by the transformed stochastic equation, whereas Table~\ref{table3} is obtained by applying the pushforward rule \eqref{eq:density_pushforward} to the Beta-sector hierarchy in Table~\ref{table1}. The hierarchy is therefore dynamical rather than purely algebraic: the lower branches arise as controlled reductions of the stochastic dynamics, so the parameters retain a direct dynamical interpretation.

\subsection{Top-level differences: \texorpdfstring{$mGB$}{mGB} versus \texorpdfstring{$mB_{\alpha}$}{mBalpha} versus \texorpdfstring{$GB$}{GB}\label{traditional}}

From the top-level densities of mGB \eqref{eq:mGB_pdf_dyn} and of GB \eqref{eq:GB_traditional}, we observe that they differ not only in normalization, but also in all three characteristic exponents, as previously discussed for mGB and mB$_\alpha$, Eqs.~\eqref{eq:mBalpha}--\eqref{eq:mGB}. Comparison of GB, mGB and mB$_\alpha$ can be summarized as follows:
\begin{align}
f_{mGB}(x)
\propto
\left(\frac{x^{\alpha}}{\beta_2}\right)^{\frac{p+\alpha -2}{\alpha}}
\left(1-\frac{x^{\alpha}}{\beta_1}\right)^{\frac{q}{\alpha}-1}
\left(1+\frac{x^{\alpha}}{\beta_2}\right)^{-\frac{p+q}{\alpha}-1} 
\nonumber \\
\quad
f_{mB_{\alpha}}(x)
\propto
\left(\frac{x^{\alpha}}{\beta_2}\right)^{\frac{\alpha p-1}{\alpha}}
\left(1-\frac{x^{\alpha}}{\beta_1}\right)^{q-1}
\left(1+\frac{x^{\alpha}}{\beta_2}\right)^{-p-q-1}
\nonumber \\
\quad
f_{GB}(x)
\propto
\left(\frac{x^{\alpha}}{\beta_2}\right)^{\frac{\alpha p-1}{\alpha}}
\left(1-\frac{x^{\alpha}}{\beta_1}\right)^{q-1}
\left(1+\frac{x^{\alpha}}{\beta_2}\right)^{-p-q}.
\label{eq:mGB_v_mBalpha_v_GB}
\end{align}
The differences between the distributions are a genuine consequence of performing the nonlinear deformation at the SDE level before deriving the stationary state. Furthermore, the difference in crossover exponent ($-p-q$) for GB versus ($-p-q-1$) for mB$_\alpha$ can be traced to the fact that mB was derived from an SDE while B was proposed ad hoc, without grounding in an underlying first-principles model -- see Eq.(15) in footnote of \cite{liu2023rethinking}. Accordingly, the traditional GB family, the dynamical-route mGB, and the distributional-route mB$_{\alpha}$ should be regarded as three distinct top-level constructions.

\subsection{Lower-branch agreement}

Although the top-level families differ, their lower branches play analogous roles. In the dynamical route, the one- and two-parameter reductions are denoted by GIGa1, GBP, GB1, GIGa, and GGa, as in Table~\ref{table2}. In the distributional route, the corresponding branches are denoted by $\text{IGa1}_{\alpha}$, $\text{BP/B2}_{\alpha}$, $\text{B}1_{\alpha}$, $\text{IGa}_{\alpha}$, and $\text{Ga}_{\alpha}$, as in Table~\ref{table3}; these are obtained branch by branch from Table~\ref{table1} through the pushforward rule \eqref{eq:density_pushforward}. Schematically,
\begin{align}
GB
&\longrightarrow
GB1,\; GB2
\longrightarrow
GGa,\; GIGa ,
\label{eq:traditional_tree_compare}
\\[4pt]
mGB
&\longrightarrow
GIGa1,\; GBP,\; GB1
\longrightarrow
GIGa,\; GGa ,
\label{eq:dynamical_tree_compare}
\\[4pt]
mB_{\alpha}
&\longrightarrow
IGa1_{\alpha},\; BP/B2_{\alpha},\; B1_{\alpha}
\longrightarrow
IGa_{\alpha},\; \mathrm{Ga}_{\alpha} .
\label{eq:distributional_tree_compare}
\end{align}
The main novelty of the present work therefore lies not in the existence of the lower branches themselves, but in the fact that a single stochastic framework unifies them and, under generalized deformation, leads to two distinct top-level constructions.

The traditional Generalized Beta family thus provides the standard algebraic hierarchy, whereas the present framework reorganizes the same landscape from the viewpoint of stochastic dynamics. Its main addition is not another parametric variant of GB, but two natural stochastic extensions of mB top level, namely the dynamical-route $\text{mGB}$ family and the distributional-route $\text{mB}_{\alpha}$ family. In this sense, $\text{mGB}$ and $\text{mB}_{\alpha}$ are best viewed not as replacements for the traditional GB family, but as stochastic extensions of it.

\section{Discussion}

The central result of this work is that the generalized extension of the modified-Beta hierarchy is \emph{not unique}. Starting from a single mean-reverting stochastic differential equation with a three-parameter stochastic term, we first obtained a complete Beta-sector hierarchy in which the modified-Beta distribution sits at the top level and controlled parameter reductions generate compact-support, heavy-tailed, and exponential-type branches. We then showed that the power transformation $y=x^\alpha$ leads to two natural generalized hierarchies: one obtained by transforming the stochastic differential equation before deriving the stationary state, and one obtained by transforming the stationary distributions afterward. These two procedures preserve the same hierarchical reduction pattern, but they
do not in general give the same generalized distributions. Their agreement should
therefore be understood at the level of branch topology and limiting distribution
classes, not as an equality of PDFs for arbitrary \(\alpha\). The generalized hierarchy is therefore route-dependent: nonlinear transformation and stationary-state reduction do not commute.

This result reorganizes the generalized-Beta landscape from the viewpoint of stochastic dynamics rather than from probability-density taxonomy alone. In the present framework, Beta-type, Beta-prime-type, Gamma-type, and inverse-Gamma-type families appear not as isolated parametric choices, but as dynamically connected stationary states generated by one underlying stochastic mechanism. The hierarchy is therefore structural rather than merely classificatory: its branches arise through controlled reductions of the same master process, and their parameters retain a direct dynamical interpretation.

This perspective also clarifies the relation to the traditional Generalized Beta family. The standard GB construct is introduced at the level of probability density functions and provides an efficient algebraic organization of related distributions. By contrast, the present framework begins with stochastic dynamics and derives the hierarchy as a hierarchy of stationary states. Its main addition is therefore not another reparameterization of the traditional GB family, but two distinct stochastic extensions of the modified-Beta top level: the dynamical-route mGB family and the distributional-route mB$_\alpha$ family. In this sense, the present construction complements the conventional GB framework by identifying a common stochastic origin for the hierarchy and by isolating what is genuinely new at the generalized top level.

The two generalized routes are useful in different ways. The dynamical route remains closer to the generating stochastic process and therefore gives a more direct interpretation of how the deformation parameter enters the stationary state. The distributional route is algebraically simpler and remains closer to the incomplete-beta structure inherited from the Beta sector, which may make it more convenient for explicit formulas and data fitting. It also remains mathematically close to the traditional GB structure. Which route is preferable depends on whether the main goal is dynamical interpretation or algebraic tractability.

More broadly, the hierarchy developed here suggests that bounded, heavy-tailed, and exponential-type distributions should often be viewed as neighboring regimes of a single stochastic construction rather than as unrelated families. This viewpoint may be useful beyond the generalized-Beta setting whenever one wants to understand how support structure, tail behavior, and limiting forms emerge from the same underlying stochastic mechanism.

Several directions remain open. A natural next step is to compare the empirical performance of the two top-level generalized families, especially in settings where endpoint behavior, tail diagnostics, or parameter stability can distinguish them. It would also be useful to study their moments, asymptotics, and fitting properties more systematically, and to ask whether similar dual generalized constructions arise in stochastic hierarchies beyond the generalized-Beta class.

\clearpage
\bibliography{mybib}

\clearpage
\appendix

\noindent\textbf{Summary of results.}
Across both the Beta sector and the two generalized routes, the workflow is: start from the specified SDE, identify the stationary distribution, and then take controlled parameter reductions to obtain the limiting families. In practice, the final outputs needed for modeling are the closed-form PDFs/CDFs and the parameter maps, together with a clear bookkeeping of which reductions connect which distributions. These are summarized in Tables~\ref{table1}--\ref{table3} and in Figs.~\ref{fig:tree1}--\ref{fig:tree3}.
\section{Compact summary of the hierarchy\label{compact}}

\begin{figure}[t]
\centering
\includegraphics[width=0.65\linewidth]{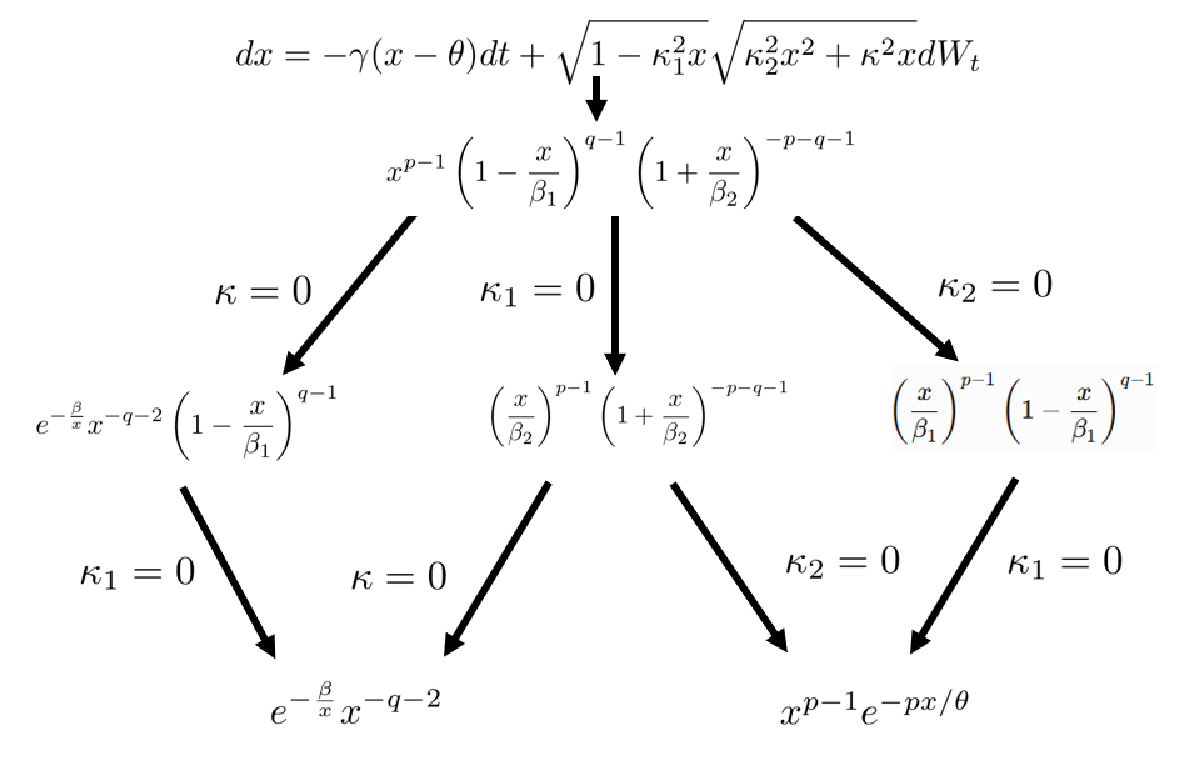}
\caption{Hierarchy in the Beta sector generated by the master SDE. Node names match the notation used throughout the text and in Table~\ref{table1}.}
\label{fig:tree1}
\end{figure}

\begin{figure}[t]
\centering
\includegraphics[width=0.65\linewidth]{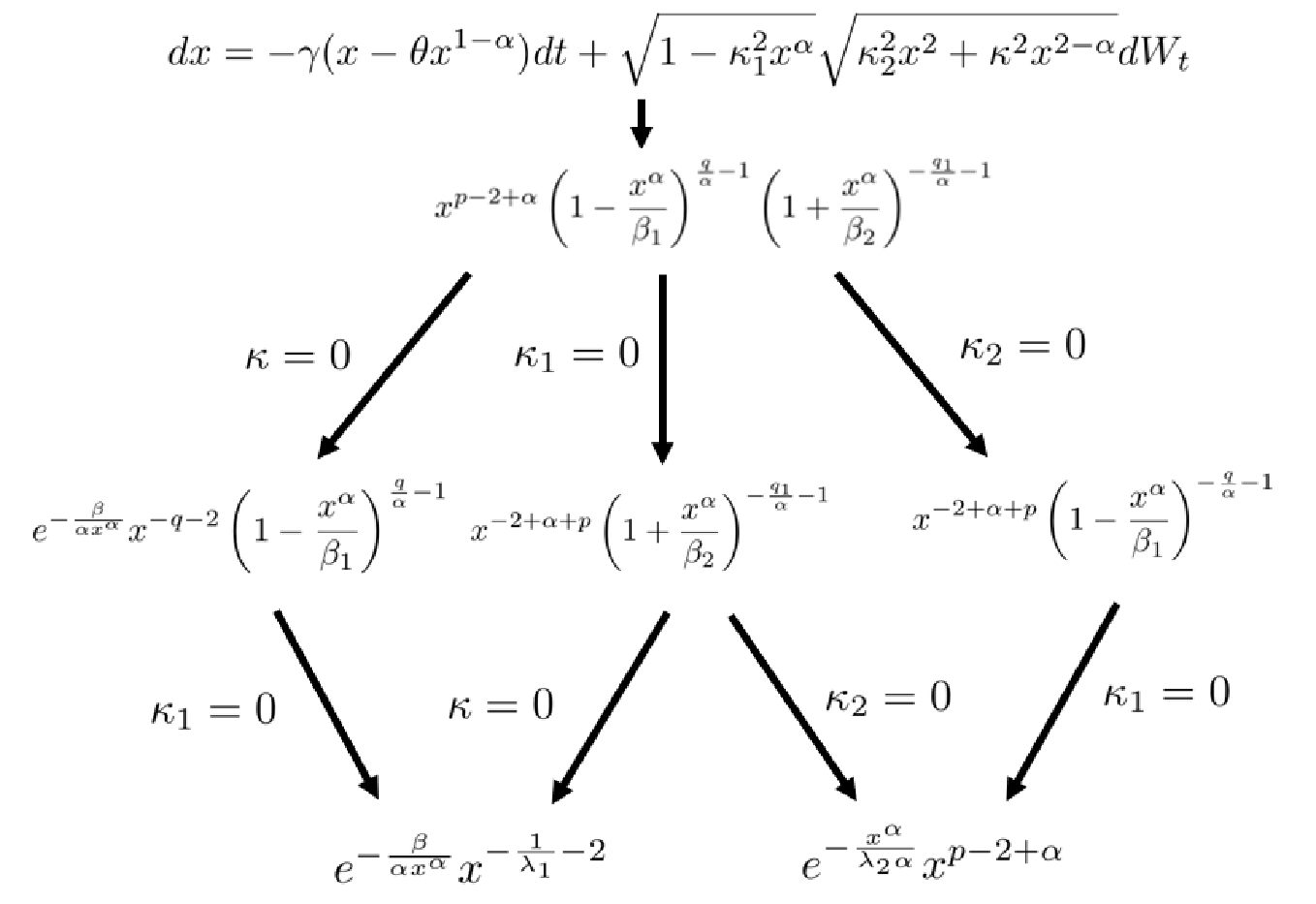}
\caption{Generalized-sector hierarchy via the two routes (distributional $mB_{\alpha}$ and dynamical $mGB$). Node names match Tables~\ref{table3} and \ref{table2}.}
\label{fig:tree2}
\end{figure}

\begin{figure}[t]
\centering
\includegraphics[width=0.75\linewidth]{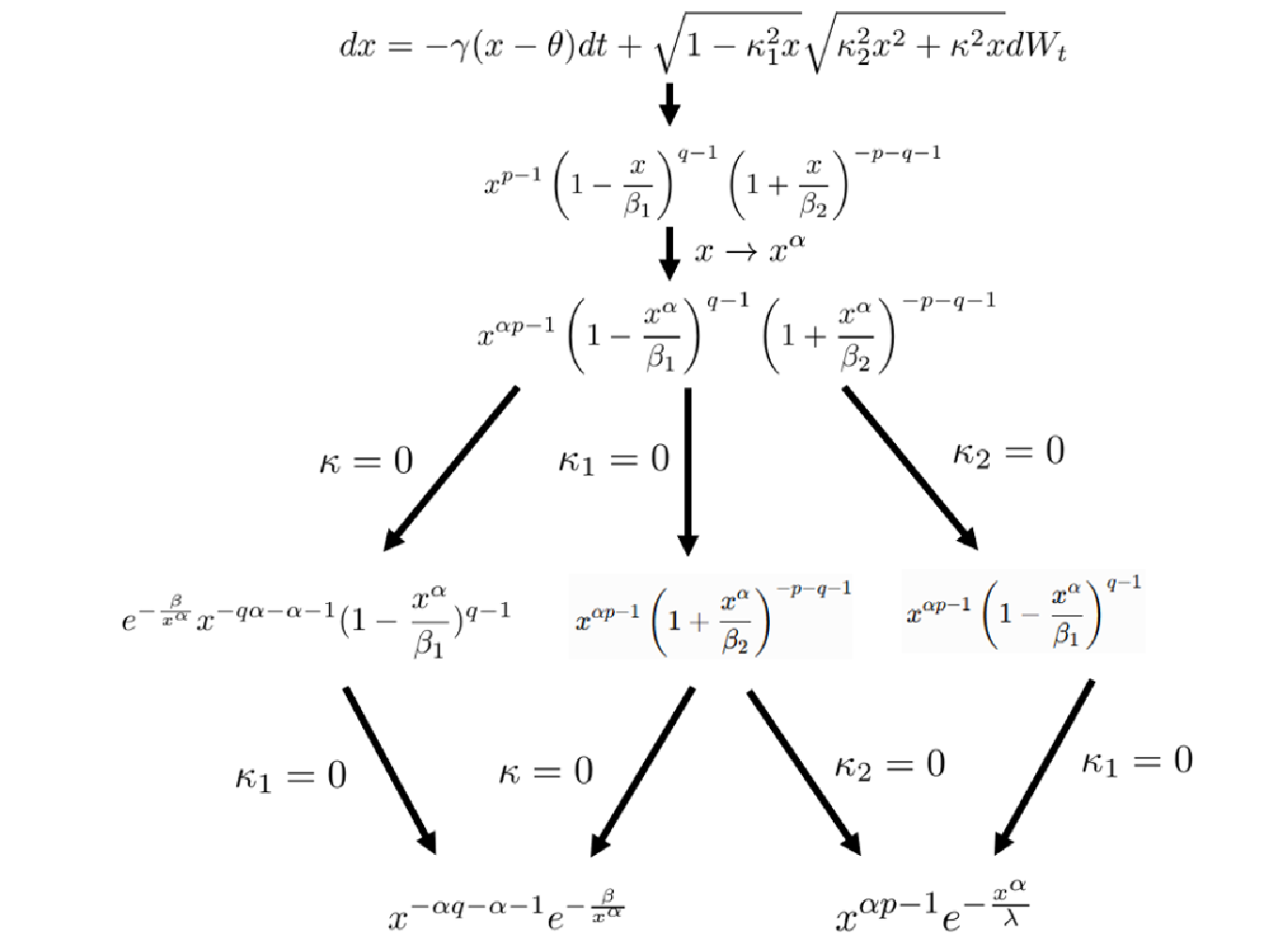}
\caption{Power-transformed hierarchy ($mB_{\alpha}$ route), shown as a compact summary figure. Node names match Table~\ref{table3}.}
\label{fig:tree3}
\end{figure}

For convenience, we collect the main branches of the hierarchy discussed in the bulk of the text. The explicit SDE definitions and closed-form PDF/CDF for each branch are given above; here we provide a compact summary of the reductions (Tables~\ref{table1}--\ref{table3p}) and the hierarchy diagrams (Figs.~\ref{fig:tree1}--\ref{fig:tree3}).

\clearpage
\begin{equation}
dx
=
-\gamma(x-\theta)\,dt
+
\sqrt{
(1-\kappa_1^2 x)
(\kappa^2 x+\kappa_2^2 x^2)
}\,dW_t ,
\end{equation}

\begin{table}[h!]
\centering
\small
\caption{mB Family}
\setlength{\tabcolsep}{2.5pt}
\renewcommand{\arraystretch}{1.25}
\begin{tabular}{|l c c c|}
\hline
SDE  & Type & PDF & CDF \\
\hline
$\kappa,\kappa_1,\kappa_2\neq0$
& mB
& $
c_{\mathrm{mB}}\left(\frac{x}{\beta_1}\right)^{p-1}
\left(1-\frac{x}{\beta_1}\right)^{q-1}
\left(1+\frac{x}{\beta_2}\right)^{-p-q-1}$
& $
I\!\left(\frac{x/\beta_1+x/\beta_2}{1+x/\beta_2};p,q\right)
{}+C_{\mathrm{mB}}
\left(\frac{1-x/\beta_1}{1+x/\beta_2}\right)^q
\left(\frac{(1+\beta_1/\beta_2)x/\beta_1}{1+x/\beta_2}\right)^p$ \\
\hline
$\kappa=0$
& IGa1
& $c_{\mathrm{IGa1}}e^{-\beta/x}
\left(\frac{x}{\beta}\right)^{-q-2}
\left(1-\frac{x}{\beta_1}\right)^{q-1}$
& $C_{\mathrm{IGa1}}\left[
\beta\Gamma\!\left(q,\frac{\beta}{x}-\frac{\beta}{\beta_1}\right)\right.
\left.{}+\beta_1\Gamma\!\left(q+1,\frac{\beta}{x}-\frac{\beta}{\beta_1}\right)
\right]$ \\
\hline
$\kappa_1=0$
& BP/B2
& $c_{\mathrm{BP}}\left(\frac{x}{\beta_2}\right)^{p-1}
\left(1+\frac{x}{\beta_2}\right)^{-p-q-1}$
& $I\!\left(\frac{x}{x+\beta_2};p,q+1\right)$ \\
\hline
$\kappa_2=0$
& B1
& $c_{\mathrm{B1}}\left(\frac{x}{\beta_1}\right)^{p-1}
\left(1-\frac{x}{\beta_1}\right)^{q-1}$
& $I\!\left(\frac{x}{\beta_1};p,q\right)$ \\
\hline
\makecell[l]{$\kappa=0$ and\\$\kappa_1=0$}
& IGa
& $c_{\mathrm{IGa}}\left(\frac{x}{\beta}\right)^{-q-2}e^{-\beta/x}$
& $C_{\mathrm{IGa}}\Gamma\!\left(q+1,\frac{\beta}{x}\right)$ \\
\hline
\makecell[l]{$\kappa_1=0$ and\\$\kappa_2=0$}
& Ga
& $c_{\mathrm{Ga}}\left(\frac{x}{\lambda}\right)^{p-1}e^{-x/\lambda}$
& $C_{\mathrm{Ga}}\gamma\!\left(p,\frac{x}{\lambda}\right)$ \\
\hline

\hline
SDE  & Type & PDF normalization & CDF normalization \\
\hline
$\kappa,\kappa_1,\kappa_2\neq0$
& mB
& $\makecell{
c_{\mathrm{mB}}=
\frac{(p+q)(1+\beta_2/\beta_1)^{p+1}(\beta_1/\beta_2)^p}
{\beta_1[(p+q)\beta_2/\beta_1+q]B(p,q)}\\
{}=\frac{(p+q)(1+\beta_1/\beta_2)^{p+1}}
{\beta_1[p+(1+\beta_1/\beta_2)q]B(p,q)}}$
& $C_{\mathrm{mB}}=
\frac{1}{B(p,q)[q+(\beta_2/\beta_1)(p+q)]}$ \\
\hline
$\kappa=0$
& IGa1
& $c_{\mathrm{IGa1}}=
\frac{\beta_1}{\beta e^{-\beta/\beta_1}(q\beta_1+\beta)\Gamma(q)}$
& $C_{\mathrm{IGa1}}=\frac{1}{(\beta+q\beta_1)\Gamma(q)}$ \\
\hline
$\kappa_1=0$
& BP/B2
& $c_{\mathrm{BP}}=\frac{1}{\beta_2B(p,q+1)}$
& --- \\
\hline
$\kappa_2=0$
& B1
& $c_{\mathrm{B1}}=\frac{1}{\beta_1B(p,q)}$
& --- \\
\hline
\makecell[l]{$\kappa=0$ and\\$\kappa_1=0$}
& IGa
& $c_{\mathrm{IGa}}=\frac{1}{\beta\Gamma(q+1)}$
& $C_{\mathrm{IGa}}=\frac{1}{\Gamma(q+1)}$ \\
\hline
\makecell[l]{$\kappa_1=0$ and\\$\kappa_2=0$}
& Ga
& $c_{\mathrm{Ga}}=\frac{1}{\lambda\Gamma(p)}$
& $C_{\mathrm{Ga}}=\frac{1}{\Gamma(p)}$ \\
\hline
\end{tabular}
\label{table1}
\end{table}

\begin{table}[h!]
\centering
\small
\caption{Parameters and Special Functions for mB Family}
\begin{tabular}{|l|}
\hline
Parameters: $\beta=\tfrac{2\gamma\theta}{\kappa_2^2}$,
$\beta_1=\tfrac{1}{\kappa_1^2}$,
$\beta_2=\tfrac{\kappa^2}{\kappa_2^2}$,
$p=\tfrac{2\gamma\theta}{\kappa^2}$,
$q=\tfrac{2\gamma(1-\kappa_1^2\theta)}{\kappa^2\kappa_1^2+\kappa_2^2}$,
$\lambda=\tfrac{\theta}{p}$ \\
\hline
$\Gamma(a,z)=\int_z^\infty t^{a-1}e^{-t}\,dt$ \quad (upper incomplete gamma function) \\
\hline
$\gamma(a,z)=\int_0^z t^{a-1}e^{-t}\,dt$ \quad (lower incomplete gamma function) \\
\hline
$I(z;p,q)=B(z;p,q)/B(p,q)$ \quad (regularized incomplete beta function) \\
\hline
\end{tabular}
\end{table}

\clearpage

\begin{equation}
dx=-\gamma (x-\theta x^{1-\alpha}) dt + \sqrt{\left(1-\kappa_1^2 x^\alpha\right)\left(\kappa^2 x^{2-\alpha}+\kappa_2^2 x^2\right)} dW_t
\end{equation}

\begin{table}[h!]
\centering
\small
\caption{mGB Family (Generalized hierarchy from the transformed SDE: dynamical route)}
\setlength{\tabcolsep}{2pt}
\renewcommand{\arraystretch}{1.28}
\begin{tabular}{|l c c c|}
\hline
SDE  & Type & PDF & CDF \\
\hline
$\kappa,\kappa_1,\kappa_2\neq0$
& mGB
& $c_{\mathrm{mGB}}
\left(\frac{x^\alpha}{\beta_2}\right)^{u-\delta}
\left(1-\frac{x^\alpha}{\beta_1}\right)^{v-1}
\left(1+\frac{x^\alpha}{\beta_2}\right)^{-w}$
& $C_{\mathrm{mGB}}
\left(\frac{x^\alpha}{\beta_2}\right)^u
F_1\!\left(u;1-v,w;1+u;
\frac{x^\alpha}{\beta_1},-\frac{x^\alpha}{\beta_2}\right)$ \\
\hline
$\kappa=0$
& GIGa1
& $c_{\mathrm{GIGa1}}e^{-\beta/(\alpha x^\alpha)}
\left(\frac{\alpha x^\alpha}{\beta}\right)^{-v-2\delta}
\left(1-\frac{x^\alpha}{\beta_1}\right)^{v-1}$
& No compact closed-form expression \\
\hline
$\kappa_1=0$
& GBP
& $c_{\mathrm{GBP}}
\left(\frac{x^\alpha}{\beta_2}\right)^{u-\delta}
\left(1+\frac{x^\alpha}{\beta_2}\right)^{-w}$
& $\makecell{C_{\mathrm{GBP}}
\left(\frac{x^\alpha}{\beta_2}\right)^u\\
{}_2F_1\!\left(w,u;1+u;-\frac{x^\alpha}{\beta_2}\right)}$ \\
\hline
$\kappa_2=0$
& GB1
& $c_{\mathrm{GB1}}
\left(\frac{x^\alpha}{\beta_1}\right)^{u-\delta}
\left(1-\frac{x^\alpha}{\beta_1}\right)^{v-1}$
& $\makecell{C_{\mathrm{GB1}}
\left(\frac{x^\alpha}{\beta_1}\right)^u\\
{}_2F_1\!\left(1-v,u;1+u;\frac{x^\alpha}{\beta_1}\right)}$ \\
\hline
\makecell[l]{$\kappa=0$ and\\$\kappa_1=0$}
& GIGa
& $c_{\mathrm{GIGa}}
\left(\frac{\alpha x^\alpha}{\beta}\right)^{-\nu-\delta}
e^{-\beta/(\alpha x^\alpha)}$
& $C_{\mathrm{GIGa}}\Gamma\!\left(\nu,\frac{\beta}{\alpha x^\alpha}\right)$ \\
\hline
\makecell[l]{$\kappa_1=0$ and\\$\kappa_2=0$}
& GGa
& $c_{\mathrm{GGa}}
\left(\frac{x^\alpha}{\alpha\lambda_2}\right)^{u-\delta}
e^{-x^\alpha/(\alpha\lambda_2)}$
& $1-C_{\mathrm{GGa}}\Gamma\!\left(u,\frac{x^\alpha}{\alpha\lambda_2}\right)$ \\
\hline

\hline
SDE  & Type & PDF normalization & CDF normalization \\
\hline
$\kappa,\kappa_1,\kappa_2\neq0$
& mGB
& $c_{\mathrm{mGB}}=
\frac{\alpha}{\beta_1^\delta}
\left(\frac{\beta_2}{\beta_1}\right)^{u-\delta}
{}\times\left[B(u,v){}_2F_1\!\left(w,u;u+v;-\frac{\beta_1}{\beta_2}\right)\right]^{-1}$
& $C_{\mathrm{mGB}}=\frac{c_{\mathrm{mGB}}\beta_2^\delta}{\alpha u}$ \\
\hline
$\kappa=0$
& GIGa1
& \multicolumn{2}{c|}{$\displaystyle c_{\mathrm{GIGa1}}=\frac{\alpha^{\delta+1}}{\beta^\delta}\left[\xi^{v+\delta}\frac{\Gamma(v)\Gamma(-v-\delta)}{\Gamma(-\delta)}{}_1F_1\!\left(1+\delta;1+v+\delta;-\xi\right)+\Gamma(v+\delta){}_1F_1\!\left(1-v;1-v-\delta;-\xi\right)\right]^{-1}$} \\
\hline
$\kappa_1=0$
& GBP
& $c_{\mathrm{GBP}}=\frac{\alpha}{\beta_2^\delta B(u,w-u)}$
& $C_{\mathrm{GBP}}=\frac{1}{uB(u,w-u)}$ \\
\hline
$\kappa_2=0$
& GB1
& $c_{\mathrm{GB1}}=\frac{\alpha}{\beta_1^\delta B(u,v)}$
& $C_{\mathrm{GB1}}=\frac{1}{uB(u,v)}$ \\
\hline
\makecell[l]{$\kappa=0$ and\\$\kappa_1=0$}
& GIGa
& $c_{\mathrm{GIGa}}=\frac{\alpha}{(\beta/\alpha)^\delta\Gamma(\nu)}$
& $C_{\mathrm{GIGa}}=\frac{1}{\Gamma(\nu)}$ \\
\hline
\makecell[l]{$\kappa_1=0$ and\\$\kappa_2=0$}
& GGa
& $c_{\mathrm{GGa}}=\frac{\alpha}{(\alpha\lambda_2)^\delta\Gamma(u)}$
& $C_{\mathrm{GGa}}=\frac{1}{\Gamma(u)}$ \\
\hline
\end{tabular}
\label{table2}
\end{table}

\begin{table}[h!]
\centering
\small
\caption{Parameters and Special Functions for mGB Family}
\setlength{\tabcolsep}{4pt}
\renewcommand{\arraystretch}{1.25}
\begin{tabular}{|l|}
\hline
\makecell[l]{Parameters inherited from the SDE: \\
$\beta=\tfrac{2\gamma\theta}{\kappa_2^2}$,
$\beta_1=\tfrac{1}{\kappa_1^2}$,
$\beta_2=\tfrac{\kappa^2}{\kappa_2^2}$,
$p=\tfrac{2\gamma\theta}{\kappa^2}$,
$q=\tfrac{2\gamma(1-\kappa_1^2\theta)}{\kappa^2\kappa_1^2+\kappa_2^2}$, 
$q_1=p+q=\tfrac{2\gamma(\kappa^2+\theta\kappa_2^2)}
{\kappa^2(\kappa^2\kappa_1^2+\kappa_2^2)}$,
$\lambda_1=\tfrac{\theta}{\beta}$,
$\lambda_2=\tfrac{\theta}{p}$.} \\
\hline
\makecell[l]{Auxiliary parameters used in Table~\ref{table2}: \\
$\delta=\tfrac{1}{\alpha}$,
$u=\tfrac{p+\alpha-1}{\alpha}$,
$v=\tfrac{q}{\alpha}$,
$w=1+\tfrac{q_1}{\alpha}$,
$\nu=\tfrac{1+1/\lambda_1}{\alpha}$,
$\xi=\tfrac{\beta}{\alpha\beta_1}$.} \\
\hline
$F_1$ = Appell hypergeometric function; \quad ${}_2F_1$ = Gauss hypergeometric function \\
\hline
${}_1F_1$ = Kummer confluent hypergeometric function \\
\hline
$\Gamma(a,z)=\int_z^\infty t^{a-1}e^{-t}\,dt$ \quad (upper incomplete gamma function) \\
\hline
\end{tabular}
\end{table}

\clearpage

\begin{table}[h!]
\centering
\small
\caption{mB$_{\alpha}$ Family (Power-transformed hierarchy: distributional route by transforming Table~\ref{table1})}
\setlength{\tabcolsep}{2pt}
\renewcommand{\arraystretch}{1.25}
\begin{tabular}{|l c c c|}
\hline
SDE & Type & PDF & CDF \\
\hline
$\kappa,\kappa_1,\kappa_2\neq0$
& mB$_\alpha$
& $c_{\mathrm{mB}}
\left(\frac{x^\alpha}{\beta_1}\right)^{p-\delta}
\left(1-\frac{x^\alpha}{\beta_1}\right)^{q-1}
\left(1+\frac{x^\alpha}{\beta_2}\right)^{-p-q-1}$
& $
I\!\left(\frac{\frac{x^\alpha}{\beta_1}+\frac{x^\alpha}{\beta_2}}
{1+\frac{x^\alpha}{\beta_2}};p,q\right)
{}+C_{\mathrm{mB}}
\left(\frac{1-\frac{x^\alpha}{\beta_1}}{1+\frac{x^\alpha}{\beta_2}}\right)^q
\left(\frac{(1+\frac{\beta_1}{\beta_2})x^\alpha/\beta_1}
{1+x^\alpha/\beta_2}\right)^p$ \\
\hline
$\kappa=0$
& IGa1$_\alpha$
& $c_{\mathrm{IGa1}}e^{-\beta/x^\alpha}
\left(\frac{x^\alpha}{\beta}\right)^{-q-1-\delta}
\left(1-\frac{x^\alpha}{\beta_1}\right)^{q-1}$
& $C_{\mathrm{IGa1}}\left[
\beta\Gamma\!\left(q,\frac{\beta}{x^\alpha}-\frac{\beta}{\beta_1}\right)\right.
\left.{}+\beta_1\Gamma\!\left(q+1,\frac{\beta}{x^\alpha}-\frac{\beta}{\beta_1}\right)
\right]$ \\
\hline
$\kappa_1=0$
& BP/B2$_\alpha$
& $c_{\mathrm{BP}}
\left(\frac{x^\alpha}{\beta_2}\right)^{p-\delta}
\left(1+\frac{x^\alpha}{\beta_2}\right)^{-p-q-1}$
& $I\!\left(\frac{x^\alpha}{x^\alpha+\beta_2};p,q+1\right)$ \\
\hline
$\kappa_2=0$
& B1$_\alpha$
& $c_{\mathrm{B1}}
\left(\frac{x^\alpha}{\beta_1}\right)^{p-\delta}
\left(1-\frac{x^\alpha}{\beta_1}\right)^{q-1}$
& $I\!\left(\frac{x^\alpha}{\beta_1};p,q\right)$ \\
\hline
\makecell[l]{$\kappa=0$ and\\$\kappa_1=0$}
& IGa$_\alpha$
& $c_{\mathrm{IGa}}
\left(\frac{x^\alpha}{\beta}\right)^{-q-1-\delta}e^{-\beta/x^\alpha}$
& $C_{\mathrm{IGa}}\Gamma\!\left(q+1,\frac{\beta}{x^\alpha}\right)$ \\
\hline
\makecell[l]{$\kappa_1=0$ and\\$\kappa_2=0$}
& Ga$_\alpha$
& $c_{\mathrm{Ga}}
\left(\frac{x^\alpha}{\lambda}\right)^{p-\delta}e^{-x^\alpha/\lambda}$
& $C_{\mathrm{Ga}}\gamma\!\left(p,\frac{x^\alpha}{\lambda}\right)$ \\
\hline

\hline
SDE & Type & PDF normalization & CDF normalization \\
\hline
$\kappa,\kappa_1,\kappa_2\neq0$
& mB$_\alpha$
& $\makecell{c_{\mathrm{mB}}=
\frac{\alpha(p+q)(1+\beta_2/\beta_1)^{p+1}(\beta_1/\beta_2)^p}
{\beta_1^\delta[(p+q)\beta_2/\beta_1+q]B(p,q)}\\
{}=\frac{\alpha(p+q)(1+\beta_1/\beta_2)^{p+1}}
{\beta_1^\delta[p+(1+\beta_1/\beta_2)q]B(p,q)}}$
& $C_{\mathrm{mB}}=
\frac{1}{B(p,q)[q+(\beta_2/\beta_1)(p+q)]}$ \\
\hline
$\kappa=0$
& IGa1$_\alpha$
& $c_{\mathrm{IGa1}}=
\frac{\alpha\beta_1}
{\beta^\delta e^{-\beta/\beta_1}(q\beta_1+\beta)\Gamma(q)}$
& $C_{\mathrm{IGa1}}=\frac{1}{(\beta+q\beta_1)\Gamma(q)}$ \\
\hline
$\kappa_1=0$
& BP/B2$_\alpha$
& $c_{\mathrm{BP}}=\frac{\alpha}{\beta_2^\delta B(p,q+1)}$
& --- \\
\hline
$\kappa_2=0$
& B1$_\alpha$
& $c_{\mathrm{B1}}=\frac{\alpha}{\beta_1^\delta B(p,q)}$
& --- \\
\hline
\makecell[l]{$\kappa=0$ and\\$\kappa_1=0$}
& IGa$_\alpha$
& $c_{\mathrm{IGa}}=\frac{\alpha}{\beta^\delta\Gamma(q+1)}$
& $C_{\mathrm{IGa}}=\frac{1}{\Gamma(q+1)}$ \\
\hline
\makecell[l]{$\kappa_1=0$ and\\$\kappa_2=0$}
& Ga$_\alpha$
& $c_{\mathrm{Ga}}=\frac{\alpha}{\lambda^\delta\Gamma(p)}$
& $C_{\mathrm{Ga}}=\frac{1}{\Gamma(p)}$ \\
\hline
\end{tabular}
\label{table3}
\end{table}

\begin{table}[h!]
\centering
\small
\caption{Parameters and Special Functions for mB$_{\alpha}$ Family}
\begin{tabular}{|l|}
\hline
Parameters: $\beta=\tfrac{2\gamma\theta}{\kappa_2^2}$,
$\beta_1=\tfrac{1}{\kappa_1^2}$,
$\beta_2=\tfrac{\kappa^2}{\kappa_2^2}$,
$p=\tfrac{2\gamma\theta}{\kappa^2}$,
$q=\tfrac{2\gamma(1-\kappa_1^2\theta)}{\kappa^2\kappa_1^2+\kappa_2^2}$,
$\lambda=\tfrac{\theta}{p}$,
$\delta=\tfrac{1}{\alpha}$ \\
\hline
$\Gamma(a,z)=\int_z^\infty t^{a-1}e^{-t}\,dt$ \quad (upper incomplete gamma function) \\
\hline
$\gamma(a,z)=\int_0^z t^{a-1}e^{-t}\,dt$ \quad (lower incomplete gamma function) \\
\hline
$I(z;p,q)=B(z;p,q)/B(p,q)$ \quad (regularized incomplete beta function) \\
\hline
\end{tabular}
\label{table3p}
\end{table}

\section{It\^o Drift Term.\label{Ito}}
By It\^o calculus a transformation to the variable $y=x^\alpha$ results in the following extra term in the r.h.s. of Eq.~\eqref{eq:mGB_sde_dyn_kappa}
\begin{equation}
\frac{1-\alpha}{2}\left(1-\kappa_1^2 x^\alpha \right) \left( \kappa^2 x^{1-\alpha} + \kappa_2^2 x \right) dt
\label{eq:ito_drift}
\end{equation}
with $\gamma$, $\kappa_1$, and $\kappa_2$ renormalized by $1/\alpha$, leading to $p$, $q$ and $q_1$ being renormalized by $\alpha$, per Eqs.~\eqref{eq:pq_map} and (\ref{eq:q1_def_dyn}). Addition of the It\^o drift term leads to the change of the exponent $\frac{p+\alpha-2}{\alpha}$ to $\frac{p-1}{\alpha}$ in Eq.~\eqref{eq:mGB_pdf_dyn}. Taking into account renormalization of $p$, $q$ and $q_1$, one recovers the pushforward hierarchy of Eq.~\eqref{eq:mBalpha_pdf_dist}.

Clearly, It\^o drift term, Eq.~\eqref{eq:ito_drift}, lacks any physical transparency and its sole role is to yield an SDE for the pushforward hierarchy. This is in contrast to its SDE counterpart
\begin{equation}
-\gamma\left( x-\theta x^{1-\alpha}\right) dt
\label{eq:relax}
\end{equation}
which identifies physically motivated relaxation process in terms of $\gamma$, $\theta$ and now $\alpha$. This makes generation of an alternative dynamical hierarchy, based directly on  Eq.~\eqref{eq:mGB_sde_dyn_kappa}, far more appealing and straightforward.

\end{document}